\documentclass[aps,prmaterials,twocolumn,superscriptaddress]{revtex4-2}

\usepackage[version=4]{mhchem}
\usepackage{gensymb}
\usepackage{xr}
\usepackage{multirow}
\usepackage{graphicx}
\usepackage{longtable}

\usepackage{erewhon}

\begin{document}


\title{Domain Walls and Defects in Ferroelectric Inorganic Halide Perovskites \ce{CsGeX3} (X = Cl, Br, I)}


\author{Kristoffer Eggestad}
\affiliation{Department of Materials Science and Engineering, NTNU - Norwegian University of Science and Technology, NO-7491 Trondheim, Norway}

\author{Benjamin A. D. Williamson}
\affiliation{Department of Materials Science and Engineering, NTNU - Norwegian University of Science and Technology, NO-7491 Trondheim, Norway}

\author{Sverre M. Selbach}
\email{selbach@ntnu.no}
\affiliation{Department of Materials Science and Engineering, NTNU - Norwegian University of Science and Technology, NO-7491 Trondheim, Norway}


\date{\today}

\begin{abstract}
Among all-inorganic halide perovskites, the only known ferroelectrics are the family of \ce{CsGeX3} (X = Cl, Br, I). Here, we study their ferroelectric domain walls (DWs) and common point defects by density functional theory (DFT) calculations and investigate the interplay between DWs and defects. The most stable defects are V$_{\text{X}}$ and V$_{\text{Cs}}$ and the former shows low migration barriers and high mobility. In contrast to oxide ferroelectrics, the affinity between point defects and DWs is negligible, reflecting the subtle structural distortions at \ce{CsGeX3} DWs. Concomitantly, the formation energies and migration energy barriers of \ce{CsGeX3} DWs are small compared to oxides, and neither V$_{\text{X}}$ nor V$_{\text{Cs}}$ pin migrating DWs. The band gap invariance across DWs and the lack of affinity towards intrinsic charged point defects imply that conducting DWs for nanoelectronics may be challenging to realise in \ce{CsGeX3}. However, shallow $p$-type defect levels and low hole effective masses suggest that high $p$-type conductivity may be achievable in nominally ferroelectric \ce{CsGeX3}. The low DW migration energy barriers and insignificant DW pinning by point defects make \ce{CsGeX3} promising materials as robust soft ferroelectrics for high-frequency switching applications with low energy dissipation.
\end{abstract}


\maketitle

\section{Introduction}
 Halide perovskites have been mostly studied for their optoelectronic and photovoltaic properties \cite{Snaith,Snaith2,doi:10.1021/nl5048779,Snaith3}. Among these, there are also ferroelectric candidates \cite{10.1063/1.4890246,doi:10.1021/nl500390f}, such as the hybrid organic-inorganic compounds with an organic A cation, e.g. MAPI (\ce{CH3NH3PbI3}) \cite{MAPI1,MAPI2,MAPI3}. Among ferroelectric halide perovskites, the family of \ce{CsGeX3} (X = Cl, Br, I) are the only known all-inorganic compounds \cite{science_adv,Christensen}. The structural ground state is rhombohedral \textit{R}3\textit{m}, making them isostructural with prototypical perovskite oxides \ce{BaTiO3} and \ce{KNbO3} \cite{PhysRevB.81.144125, Cohen1992, PhysRev.76.1221, 10.1063/1.1625080, PhysRev.93.672, Maeder2004}, and different from MAPI which is typically reported as tetragonal at room temperature \cite{doi:10.1021/acs.jpca.5b09884}. They possess \textit{$T_C\,$}s of 155$\degree$C, 238$\degree$C, and 277$\degree$C  for X = Cl, Br, I \cite{https://doi.org/10.1002/zaac.19875450217}, respectively, and polarisations \cite{science_adv} within $10-20$ $\micro\text{C}/\text{cm}^2$. Most studies hitherto of \ce{CsGeX3} have emphasised photovoltaic properties, including several \textit{ab initio} studies of the electronic structure \cite{Li-Chuan, doi:10.1021/acs.jpcc.8b00226, BOUHMAIDI2022e00663, doi:10.1021/acs.jpcc.4c06280, AYALEW2025e01043, ELAKKEL2024115721}. \ce{CsGeCl3} has also shown promise as a $p$-type conductor, similar to \ce{Sn(II)O}, where traditionally localised deep lying holes situated on O\,$2p$ (in \ce{Sn(IV)O2}, an $n$-type conductor) are now delocalised thanks to covalency between O\,$2p$ and Ge\,$4s$, lowering the effective mass of holes and the ionisation potential allowing for low formation energy shallow defects such as $V_{\text{Cs}}$ \cite{doi:10.1021/cm401343a, Huang_2014}.
The related \ce{CsSnX3} and \ce{CsPbX3} also possess B-site lone pairs, but here tensile strain is needed to stabilise polarisation \cite{doi:10.1021/acs.chemmater.3c02201}. 

 Conductive ferroelectric DWs were first reported in La-doped \ce{BiFeO3} \cite{Seidel2009}, but have later been shown in a multitude of materials systems \cite{Meier2022, Nataf2020, https://doi.org/10.1002/adfm.201201174}. DWs in ferroelectric and ferroelastic oxides are known to become conducting \cite{Meier2022, 10.1063/5.0009185} from either intrinsic effects such as band bending at neutral or charged DWs \cite{Seidel2009, PhysRevMaterials.2.114405, CDW_LiNbO3, Meier2012}, or by extrinsic effects such as electronic charge compensation of charged point defects accumulating at the DWs \cite{Rojac2017, Schaab2018, https://doi.org/10.1002/adma.201102254}. The space group of \ce{CsGeX3}, \textit{R}3\textit{m}, allows 71$\degree$, 109$\degree$ and 180$\degree$ ferroelectric DWs. The 71$\degree$ and 109$\degree$ DWs must also be ferroelastic as the strain state changes across these walls. Our study of DWs in \ce{CsGeX3} is also motivated by the high charge carrier mobility \cite{doi:10.1021/ic401215x, doi:10.1021/acs.jpclett.6b02800}, defect tolerance \cite{doi:10.1126/science.aam7093, doi:10.1021/acs.jpclett.6b02800, doi:10.1021/acsenergylett.1c02027, Ganose2022, Ye2024} and intermediate band semiconductivity \cite{PhysRevB.53.12545, Tang_2000}, all implying that different physical DW properties compared to oxides are anticipated. 
 
 Here, we use density functional theory (DFT) calculations to study DWs and point defects in ferroelectric \ce{CsGeX3} (X = Cl, Br, I). We calculate electronic structure, DW mobility, point defect formation energies and migration barriers, and interactions between DWs and point defects. \ce{CsGeX3} exhibit significantly different DW energies, mobilities, and electronic structures compared to corresponding oxides. High DW mobility, point defect insensitivity, and propensity to $p$-type doping give the all-inorganic halide perovskites a distinctly complementary technological potential compared to oxide ferroelectrics.
\section{Computational Details}
 Point defects and DWs in \ce{CsGeX3} have been investigated using density functional theory (DFT) within the \texttt{VASP} code \cite{vasp1, vasp2, vasp3}. Interactions between core and valence electrons (Cs: (4s$^{2}$, 4p$^{6}$, 5s$^1$), Ge: (3d$^{10}$, 4s$^2$, 4p$^{2}$), Cl: (3s$^2$, 3p$^5$), Br: (4s$^2$, 4p$^5$), I: (5s$^2$, 5p$^5$)) were treated using the projector-augmented-wave method (PAW) \cite{paw1, PAW}. The HSE06 functional \cite{HSE06} (Heyd-Scuseria-Ernzerhof screened hybrid functional, with mixing parameter $\alpha=0.25$ and screening parameter $\omega=0.11$ bohr$^{-1}$) was used for all bulk structure calculations, as it yields lattice parameters in close agreement with experimental values (see Table \ref{tab:lattice_comp}) and is known to provide accurate descriptions of electronic structure and carrier localisation \cite{doi:10.1021/acs.jpclett.6b01807, B812838C}.  Due to the high computational cost of HSE06, the PBEsol \cite{PBEsol} functional (Perdew-Burke-Ernzerhof revised for solids) was used to investigate defect mobilities and domain walls. A plane-wave energy cutoff of 400 eV was used for all calculations. The effects of spin-orbit coupling (SOC) were studied for all three halide systems, however, only minor changes in the electronic structure of \ce{CsGeI3} were observed (see SI Figure S5), and thus SOC was omitted in subsequent calculations to reduce computational cost.

Electronic density of states (DOS) were calculated using primitive cells containing 5 atoms. Geometries and atomic positions were optimised until all forces on all ions were less than $1\times10^{-4}\,\text{eV}\,Å^{-1}$ using $8\times8\times8$ $\Gamma$-centred \textit{k}-point meshes. \textit{k}-points along high-symmetry paths, for electronic band structure calculations, were generated using the \texttt{SUMO} program \cite{Ganose2018}. 

Polarisation was estimated using the point charge model, with formal charges (FC) and Born effective charges (BEC), and the Berry phase method. BEC were calculated using finite differences and tabulated in SI Table S6. The finite differences method also gives elastic, piezoelectric and dielectric tensors, presented in SI Table S6, S7, S8 and S9. Similar properties have been estimated using density functional perturbation theory (DFPT) with the PBEsol functional and are displayed in SI Table S11, S12, S13 and S14.
%
 Intrinsic point defect formation energies were calculated using \cite{Freysoldt}
 \begin{equation}
 \label{eq:defect_form}
 \begin{split}
  \Delta H_{F}(D,q) =  & E_{\text{D}}(q) - E_{\text{H}} +\sum_i n_i (E_i+\mu_i) \\
                            & + q (E_{\text{F}} + E_{\text{vbm}} + E_{\text{corr}}^{\text{pot}})+ E_{\text{corr}}^{\text{IC}}+ E_{\text{corr}}^{\text{BF}}
 \end{split}
 \end{equation} 
 where $E_{\text{H}}$ is the reference energy of the pristine host cell. $E_{\text{D}}(q)$ is the energy of the supercell including the defect, D, with charge state $q$. $E_i$ and $\mu_i$ are the reference energy and the chemical potential of element $i$, respectively, and $n_i$ is the number of element $i$ removed from the system. $E_{\text{F}}$ is the Fermi level for which to evaluate the defect formation energies, while $E_{\text{VBM}}$ refers this energy to the energy of the valence band maximum in the DFT calculations. Due to the limited supercell, three correction schemes, $E_{\text{corr}}^{\text{pot}}$, $E_{\text{corr}}^{\text{IC}}$ and $E_{\text{corr}}^{BF}$, are employed to correct for finite size effects. The potential alignment correction, $E_{\text{corr}}^{\text{pot}}$, aligns the potential of the defect cell to that of the defect free host cell \cite{PAC}, and the image charge correction, $E_{\text{corr}}^{\text{IC}}$, corrects for the interactions between the defect and the charge of its periodic image \cite{PAC,IC}. Lastly, the band filling correction, $E_{\text{corr}}^{BF}$, corrects for the unphysical filling of holes and electrons in the valence and conduction band, respectively \cite{PAC,doi:10.1063/1.1682673}.

 The thermodynamic stability regions of \ce{CsGeCl3} and \ce{CsGeI3}, shown in SI Figure S6 and S7, are calculated by evaluating formation energies of all competing phases using the Chemical Potential Limits Analysis Program (\texttt{CPLAP}) \cite{cplap}. To show the full defect chemistry of \ce{CsGeCl3} and \ce{CsGeI3}, defect formation energies, shown in Figure \ref{fig:TL_Cl_I} a and b, are calculated using three sets of chemical potentials for both materials, indicated in SI Figure S6 and S7.

 The mobility of V$_{\text{Cs}}$ and V$_{\text{X}}$ were investigated using the climbing image nudged elastic bands method \cite{NEB1,NEB2} (ci-NEB) with 5 images between the initial and final image. The migration barriers were calculated in both supercells with and without charge compensation. Two different migration paths for V$_{\text{X}}$, displayed in SI Figure S9, were investigated.
%
 
 Lattice parameters for 71$\degree$, 109$\degree$ and 180$\degree$ DW cells for \ce{CsGeX3} were calculated from the equations displayed in SI Table S15 using the PBEsol optimised lattice parameters. To avoid DW-DW interactions, a-parameters of $\sim$ 75 Å were used. $\Gamma$-centred \textit{k}-point meshes of $1\times4\times6$, $1\times4\times4$, $1\times6\times4$ and $1\times6\times4$ were used for the for 71$\degree$, 109$\degree$ and 180$\degree$ Y- and X-type DW supercells, respectively. The atomic positions were relaxed until all forces on all ions were less than 0.001 eV Å$^{-1}$. DW supercells, for \ce{CsGeCl3}, with DFT optimised atomic positions are shown in SI Figure S11. 

 Investigations of defect segregation enthalpies were investigated by calculating the defect formation energies as a function of distance from a 71$\degree$ DW in \ce{CsGeCl3}. To avoid defect-defect interactions the cross section area of the DW supercells was increased by an $1\times2\times3$ expansion and a $\Gamma$-centred \textit{k}-point mesh of $1\times2\times2$ was used.

 The mobility of DWs was investigated using the ci-NEB method with 5 images. A strict force criterion of 0.001 eV Å$^{-1}$ was used for optimising the atomic positions of the start and end images due to the flat energy landscape around the DW equilibrium position. ci-NEB calculations were performed with a force criterion of 0.01 eV Å$^{-1}$. Point defect pinning was studied for 71$\degree$ DWs in \ce{CsGeCl3} by calculating migration barriers across point defects in DW supercells. The supercells used are described in SI Table S16 and Figure S17 and a $\Gamma$-centred \textit{k}-point mesh of $1\times2\times2$ and a force criterion of 0.01 eV Å$^{-1}$ was used for structure optimisation. 
\section{Results}
\begin{figure}[htbp]
    \centering
    \includegraphics[width=0.9\linewidth]{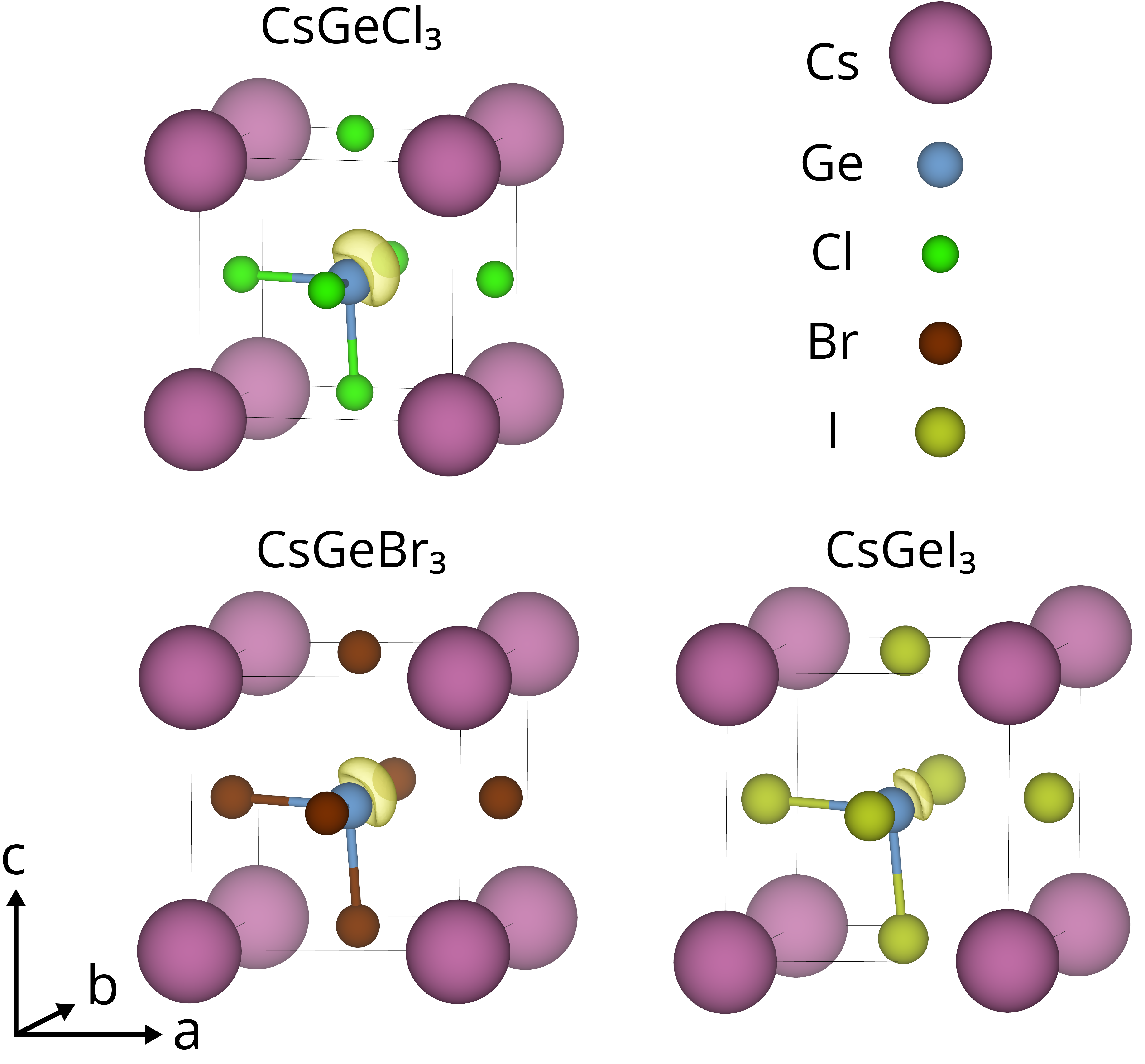}
    \caption{DFT optimised primitive structures of \ce{CsGeCl3}, \ce{CsGeBr3} and \ce{CsGeI3} visualised together with their respective lone pairs. The lone pairs are extracted from the partial charge density of the top valence bands and shown with an isosurface level of 0.02.}
    \label{fig:lone_pairs}
\end{figure}

The structural properties of the relaxed primitive cells, see Figure \ref{fig:lone_pairs}, were calculated and are presented in Table \ref{tab:lattice_comp}. Our calculated lattice parameters show good agreement with experimental values reported in the literature \cite{Christensen,https://doi.org/10.1002/zaac.19875450217,SCHWARZ199520,LIN200828}. The HSE06 functional gives excellent agreement with experiments, with a subtle overestimation, as expected due to thermal expansion and related effects.
\begin{table}[htbp]
    \centering
    \caption{Calculated and experimental lattice parameters and Ge-X bond lengths for the primitive unit cell of \ce{CsGeCl3}, \ce{CsGeBr3} and \ce{CsGeI3}. The bonds described are the bonds shown in Figure \ref{fig:lone_pairs}.}
    \begin{ruledtabular}
    \begin{tabular}{cccccc}
         & & PBE & PBEsol & HSE06 & Exp. \\
         \colrule
        \multirow{5}{*}{\ce{CsGeCl3}} & \multirow{2}{*}{a,b,c (Å)} & \multirow{2}{*}{5.519} & \multirow{2}{*}{5.294} & \multirow{2}{*}{5.513} & 5.444\cite{Christensen}, 5.434\cite{https://doi.org/10.1002/zaac.19875450217}, \\
         & & & & & 5.391\cite{LIN200828} \\
         & \multirow{2}{*}{$\alpha$,$\beta$,$\gamma$ ($\degree$)} & \multirow{2}{*}{89.00} & \multirow{2}{*}{89.76} & \multirow{2}{*}{89.49} & 89.63\cite{Christensen}, 89.72\cite{https://doi.org/10.1002/zaac.19875450217}, \\
         & & & & & 89.71\cite{LIN200828} \\
         & Ge-Cl (Å) & 2.41 & 2.42 & 2.37 & 2.35 \cite{https://doi.org/10.1002/zaac.19875450217}\\
        \colrule
        \multirow{5}{*}{\ce{CsGeBr3}} & \multirow{2}{*}{a,b,c (Å)} & \multirow{2}{*}{5.744} & \multirow{2}{*}{5.504} & \multirow{2}{*}{5.731} & 5.636\cite{SCHWARZ199520,https://doi.org/10.1002/zaac.19875450217}, \\
        & & & & & 5.638\cite{LIN200828} \\
         & \multirow{2}{*}{$\alpha$,$\beta$,$\gamma$ ($\degree$)} & \multirow{2}{*}{88.45} & \multirow{2}{*}{89.53} & \multirow{2}{*}{88.46} & 88.74\cite{SCHWARZ199520,https://doi.org/10.1002/zaac.19875450217}, \\
         & & & & & 88.73\cite{LIN200828} \\
          & Ge-Br (Å) & 2.58 & 2.60 & 2.54 & 2.53 \cite{https://doi.org/10.1002/zaac.19875450217}\\
        \colrule
        \multirow{3}{*}{\ce{CsGeI3}} & a,b,c (Å) & 6.108 & 5.860 & 6.081 & 5.983 \cite{https://doi.org/10.1002/zaac.19875450217} \\
         & $\alpha$,$\beta$,$\gamma$ ($\degree$) & 88.25 & 89.49 & 88.14 & 88.61 \cite{https://doi.org/10.1002/zaac.19875450217} \\
          & Ge-I (Å) & 2.79 & 2.80 & 2.75 & 2.75 \cite{https://doi.org/10.1002/zaac.19875450217}\\
    \end{tabular}
    \end{ruledtabular}
    \label{tab:lattice_comp}
\end{table}
Estimated polarisation using formal charges (FC), Born effective charges (BEC) and Berry phase calculations are shown in Table \ref{tab:gen} and show a decreasing polarisation from \ce{CsGeCl3} to \ce{CsGeI3}. The Berry phase method suggests significantly larger polarisations than using BEC and the point charge model, indicating significant contribution from electrons. Our values are relatively close to what has been reported previously using the PBE functional \cite{science_adv}.
\begin{table}[htbp]
    \caption{Tolerance factor, polarisation, band gaps and effective masses of holes and electrons calculated for \ce{CsGeX3} (X = Cl, Br, I) using the HSE06 functional. Born effective charges are calculated using the finite differences method.}
    \begin{ruledtabular}
    \begin{tabular}{cccc}
        & \ce{CsGeCl3} & \ce{CsGeBr3} & \ce{CsGeI3}\\
        \colrule
        \textit{t} & 1.027 & 1.009 & 0.985 \\
        P$_{FC}$ ($\micro\text{C}/\text{cm}^2$) & 8.5 & 7.6 & 6.9 \\
        P$_{BEC}$ ($\micro\text{C}/\text{cm}^2$) & 10.84 & 10.47 & 9.88 \\ 
        P$_{Berry}$ ($\micro\text{C}/\text{cm}^2$) & 24.3 & 21.3 & 21.0 \\
        E$_g$ (eV) & 3.015 & 2.068 & 1.438 \\
        $m^*_h$ & 0.325 & 0.189 & 0.138 \\
        $m^*_e$ & 0.438 & 0.187 & 0.134
    \end{tabular}
    \end{ruledtabular}
    \label{tab:gen}
\end{table}
\subsection{Electronic Structure}
\begin{figure}[htbp]
    \centering
    \includegraphics[width=0.9\columnwidth]{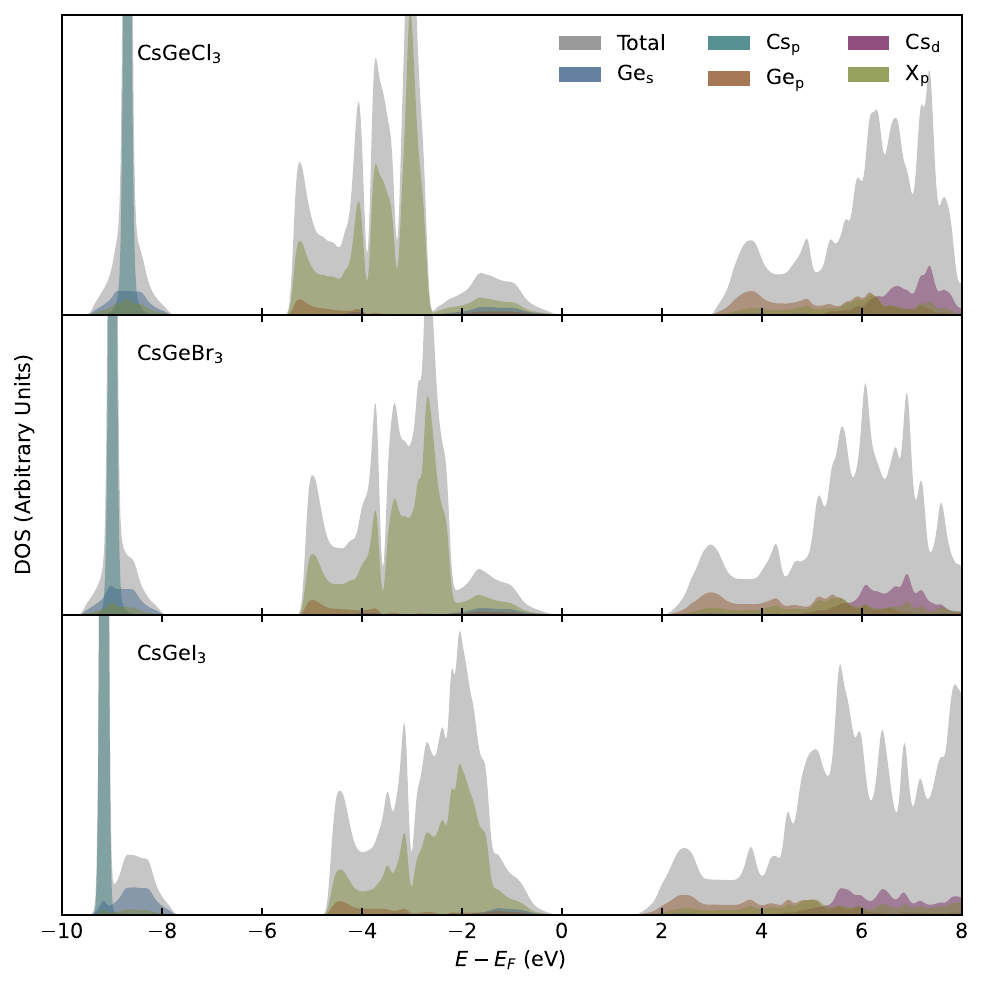}
    \caption{Electronic density of states for \ce{CsGeCl3}, \ce{CsGeBr3} and \ce{CsGeI3} calculated using the HSE06 functional. Cs\,$p$, Cs\,$d$, Ge\,$s$, Ge\,$p$ and X\,$p$ (halogen-p) are shown and coloured in turquoise, purple, blue, orange and green, respectively. Other orbitals do not contribute significantly and are thus omitted.}
    \label{fig:dos_all}
\end{figure}

 Partial and total electronic density of states (DOS) are calculated for \ce{CsGeX3} and displayed in Figure \ref{fig:dos_all}. The conduction band minimum (CBM) consist mainly of \ce{Ge}\,$p$ states with some covalency with \ce{X}\,$p$. The valence band maximum (VBM) show mixing of \ce{Ge}\,$s$, \ce{Ge}\,$p$ and \ce{X}\,$p$. The Ge lone pair can be seen at the top of the valence band as hybridisation between \ce{Ge}\,$s$ and \ce{Ge}\,$p$. The lone pair causes an off-centring of the Ge ions and thus non-centrosymmetry and ferroelectricity in these halide perovskites. 

 The strength (stereochemical activity) of the lone pair can be determined by the overlap between \ce{Ge}\,$s$ and \ce{X}\,$p$ as good overlap results in mainly non-bonding \ce{Ge}\,$s$ states that are high enough in energy for hybridisation with \ce{Ge}\,$p$. The overlap between \ce{Ge}\,$s$ and \ce{X}\,$p$ and the lone pair are shown in detail in SI Figure S4. The degree of covalency between \ce{Ge}\,$s$ and \ce{X}\,$p$ has been quantified by calculating the \ce{X}\,$p$ contribution to the covalency between \ce{Ge}\,$s$ and \ce{X}\,$p$ at around $-9$\,eV in a similar fashion as what was done by Payne et al. for a series of post-transition metal oxides \cite{PhysRevLett.96.157403,C1CS15098G}. Calculated \ce{X}\,$p$ contribution of $0.34$, $0.25$ and $0.17$ for \ce{CsGeCl3}, \ce{CsGeBr3} and \ce{CsGeI3}, respectively, predicts stronger mixing and thus stronger lone pairs going from \ce{CsGeI3} to \ce{CsGeCl3}. The energy of \ce{X}\,$p$ increases down the periodic table leading to less mixing of \ce{Ge}\,$s$ and \ce{X}\,$p$, therefore weaker hybridisation between \ce{Ge}\,$s$ and \ce{Ge}\,$p$, and a weaker lone pair. The evaluated lone pair strength correlates nicely with decreasing band gaps and polarisation from \ce{CsGeCl3} to \ce{CsGeI3}. DOS-es calculated using the PBEsol functional show very similar trends and are displayed in SI Figure S3. To further illustrate the nature of the lone pairs, Figure \ref{fig:lone_pairs} show the real-space projections of the charge density at the top of the valence band.

 Calculated band structures show direct band gaps at Z of $\sim$ 3.0, 2.1 and 1.4 eV, slightly lower than experimental values reported for \ce{CsGeCl3} and \ce{CsGeBr3} \cite{PhysRevB.53.12545}. From \ce{CsGeCl3} to \ce{CsGeI3}, both the lattice parameters and Ge-X bond lengths increase, which is inversely correlated with the observed reduction in band gap. Moreover, calculated band structures show a large degree of curvature at the band edges and display large band gaps of around 4 eV at the $\Gamma$ point demonstrating the importance of the \textit{k}-point sampling \cite{doi:10.1021/acsomega.2c02088}. Calculated effective masses, displayed in Table \ref{tab:gen}, show exceptionally low effective masses. According to the classification by Davies et al. \cite{doi:10.1021/acs.jpclett.9b03398}, which characterises the degree of polaronic behaviour, \ce{CsGeCl3} lies on the border between type I and II, while \ce{CsGeBr3} and \ce{CsGeI3} fall firmly within type I. Therefore, very high electron and hole mobilities are expected for \ce{CsGeBr3} and \ce{CsGeI3}, whereas some degree of polaronic behaviour is expected in \ce{CsGeCl3}. The PBEsol functional gives relatively similar band structures, see SI Figure S3, albeit significantly lower band gaps. The bottom two bands of the CB are almost degenerate at the F and Z points in \ce{CsGeCl3}, but less so in \ce{CsGeBr3} and \ce{CsGeI3}. The localised bands at around -9 eV are Cs\,$p$ as seen in the DOS in Figure \ref{fig:dos_all}.
\begin{figure*}[htbp]
    \centering
    \includegraphics[width=0.9\textwidth]{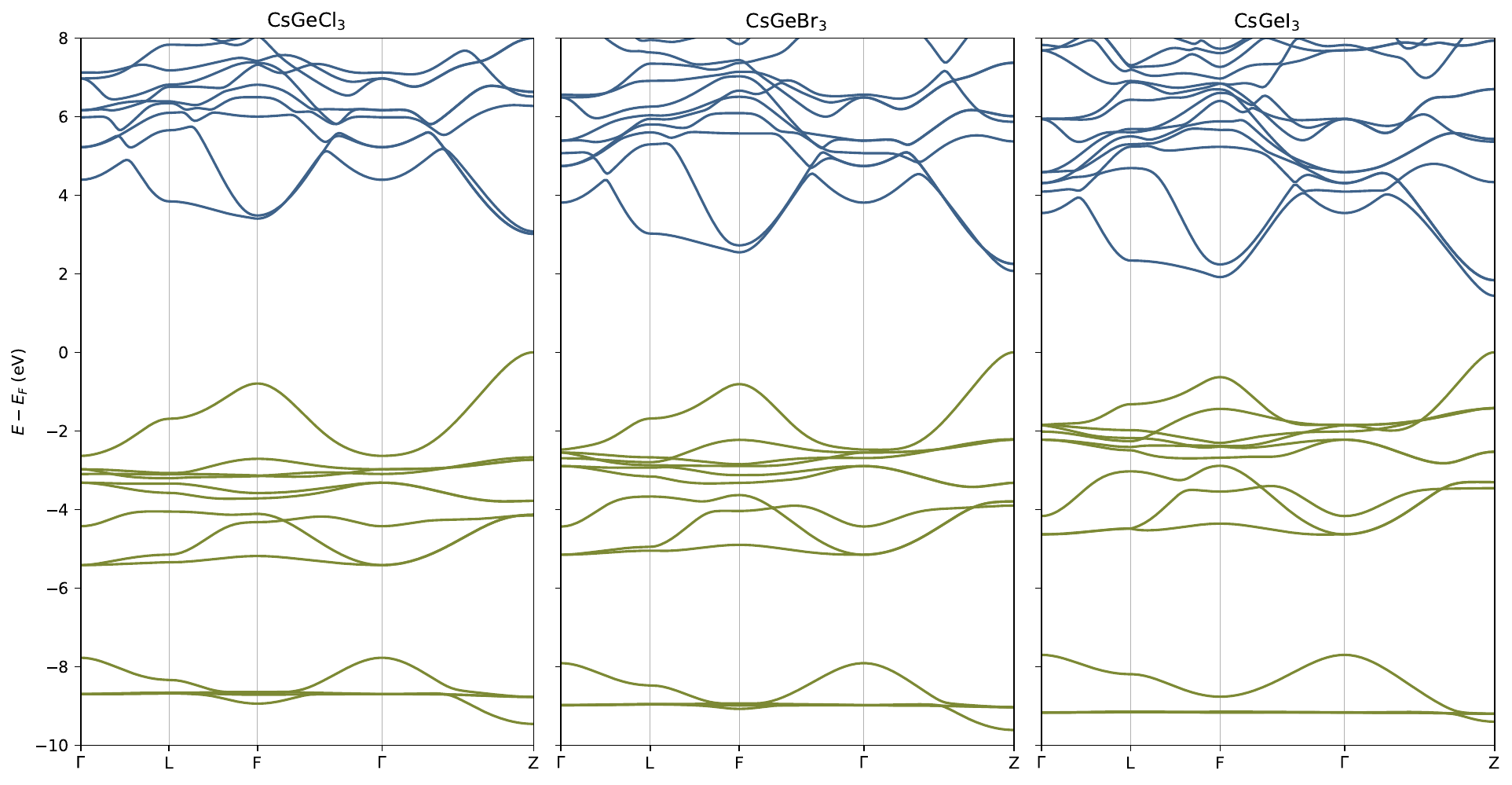}
    \caption{Electronic band structures for \ce{CsGeCl3}, \ce{CsGeBr3} and \ce{CsGeI3} calculated using the HSE06 functional. The valence and conduction band are coloured in green and blue, respectively.}
    \label{fig:band_all}
\end{figure*}
\subsection{Intrinsic Defect Thermodynamics}
\begin{figure*}[htbp]
    \centering
    \includegraphics[width=\textwidth]{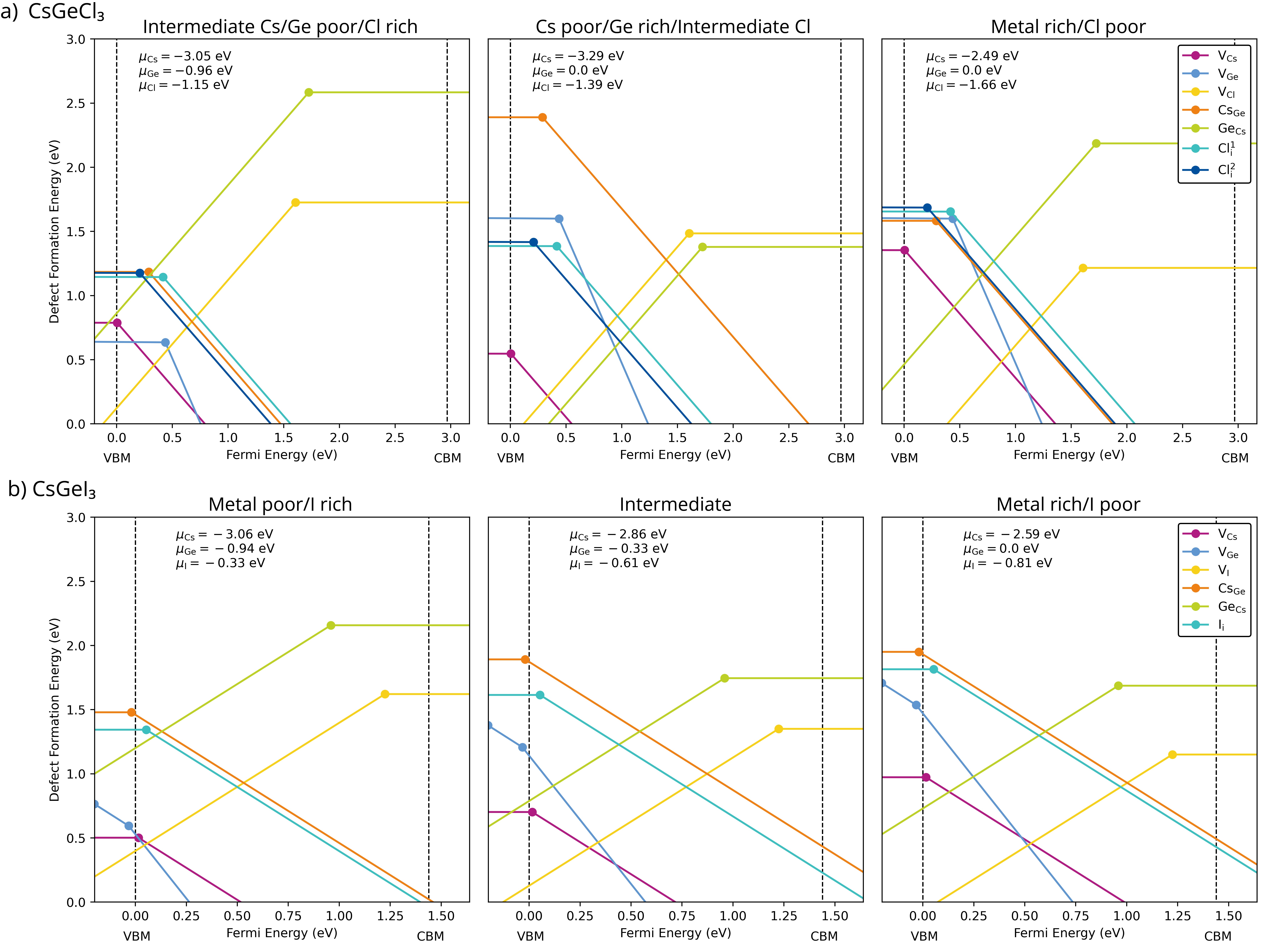}
    \caption{Thermodynamic transition level diagrams for intrinsic defects in bulk \ce{CsGeCl3} (a) and \ce{CsGeI3} (b) calculated using the HSE06 functional. The three panels for each compound show the transition levels at three different chemical environments indicated in the stability window shown in SI Figure S6 and S7. The dashed lines at 0~eV, and $\sim3$ and $\sim1.5$~eV display the VBM and CBM, respectively.}
    \label{fig:TL_Cl_I}
\end{figure*}

 Defect formation energies as a function of the Fermi energy for \ce{CsGeCl3} and \ce{CsGeI3}, calculated using the HSE06 functional, are displayed in Figure \ref{fig:TL_Cl_I} a and b. Each figure includes three panels showing the formation energies at three different parts of the stability window, indicated in SI Figure S6 and S7. Hybrid defect calculations were not performed for \ce{CsGeBr3}, as this compound is expected to exhibit intermediate properties between \ce{CsGeCl3} and \ce{CsGeI3}.

 Throughout the stability region, the defects with the lowest formation energies are V$_\text{Cs}$, V$_\text{Ge}$, and V$_\text{X}$. Halide interstitials show relatively high formation energies, and therefore, V$_\text{X}$ are expected to be the dominant halide defects. Moreover, cation antisites also show high formation energies, due to the difference in charge and size (0.73 Å, 1.88 Å\cite{Shannon}) between Cs and Ge ions.
 
 All $n$-type defects show deep transition levels while several $p$-type defects display shallow or resonant defect levels. V$_\text{Cs}$ show shallow defect levels (-1/0) of 0.004 and 0.015 eV above the VBM in \ce{CsGeCl3} and \ce{CsGeI3}, respectively, while V$_\text{Ge}$ is deep in \ce{CsGeCl3} and resonant in \ce{CsGeI3}. At the halogen rich limit (left panels), the Fermi level is pinned at approximately 0.4 and 0.07 eV above the VBM for \ce{CsGeCl3} and \ce{CsGeI3}, respectively. Both materials show $p$-type dopeability at this part of the stability region. Towards the halogen poor edge, the Fermi level shifts towards the middle of the band gap and less $p$-type character is observed.

 Point defects and $p$-type conductivity in \ce{CsGeCl3} have previously been investigated, using DFT with the PBE functional, by Huang et al.\cite{Huang_2014}. Despite the different functional, our calculations show similar results to what they report, but as expected, the HSE06 functional gives deeper transition levels as HSE06 is known to better localise holes and electrons. Using the HSE06 functional also results in subtly larger formation energies on average.

\begin{figure}[htbp]
    \centering
    \includegraphics[width=0.9\columnwidth]{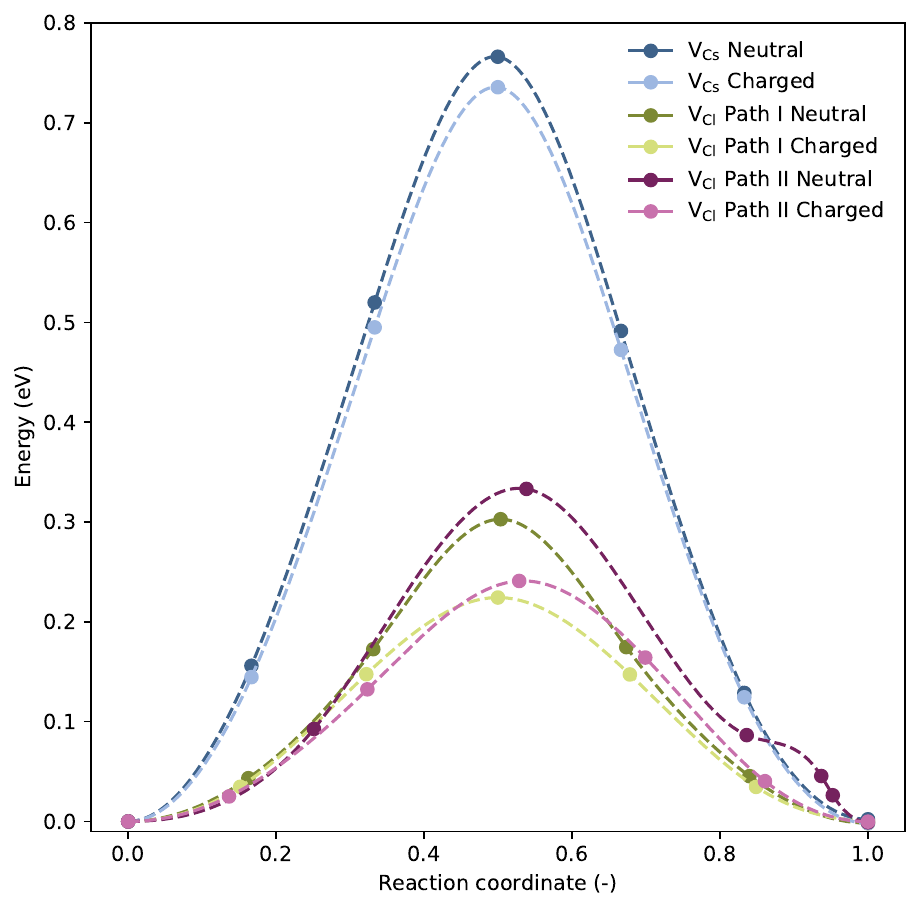}
    \caption{Migration barriers of V$_{\text{Cs}}$ and V$_{\text{Cl}}$ in bulk \ce{CsGeCl3} in charged and neutral cells calculated using the PBEsol functional.}
    \label{fig:defect_mig}
\end{figure}

 Calculated migration barriers for V$_\text{Cs}$ and V$_\text{Cl}$ in both neutral and charged supercells of \ce{CsGeCl3} are displayed in Figure \ref{fig:defect_mig}. V$_\text{Cs}$ show relatively low migration barriers of $\sim$ 0.75 eV, while V$_\text{Cl}$ display very low migration barriers comparable with V$_\text{Li}$ and V$_\text{Na}$ in the best solid state batteries \cite{Deng2016,https://doi.org/10.1002/anie.200701144}. In the charged cell, without any charge compensation, the migration barriers are $\sim$ 0.22 eV and $\sim$ 0.24 eV, and in the charged cell, compensated by one extra electron, $\sim$ 0.30 eV and $\sim$ 0.33 eV for path I and II, respectively. Path I is between two Cl sites with their short Ge-Cl bonds to the same Ge atom while path II is between two Cl sites with their short bonds to two different Ge atoms. Both paths are displayed in SI Figure S9. Migration of charge compensated V$_\text{Cl}$ through path II shows an additional small barrier due to polaron hopping.
 
 V$_\text{Cs}$ and V$_\text{I}$ in \ce{CsGeI3} show very similar migration barriers to what is shown for \ce{CsGeCl3}, see SI Figure S8. Mobile ions are known as one of the main contributors to the degradation of halide perovskite solar cells \cite{doi:10.1126/science.aam6323,doi:10.1021/jacs.5b08535}, however, for instance, calculated migration barrier for V$_{\text{I}}$ in methylammonium lead iodide (\ce{CH3NH3PbI3}) of 0.6 eV \cite{Eames2015} is significantly higher than what is estimated in this work.

\subsection{Domain Walls}
\begin{figure}[ht]
    \centering
    \includegraphics[width=0.9\linewidth]{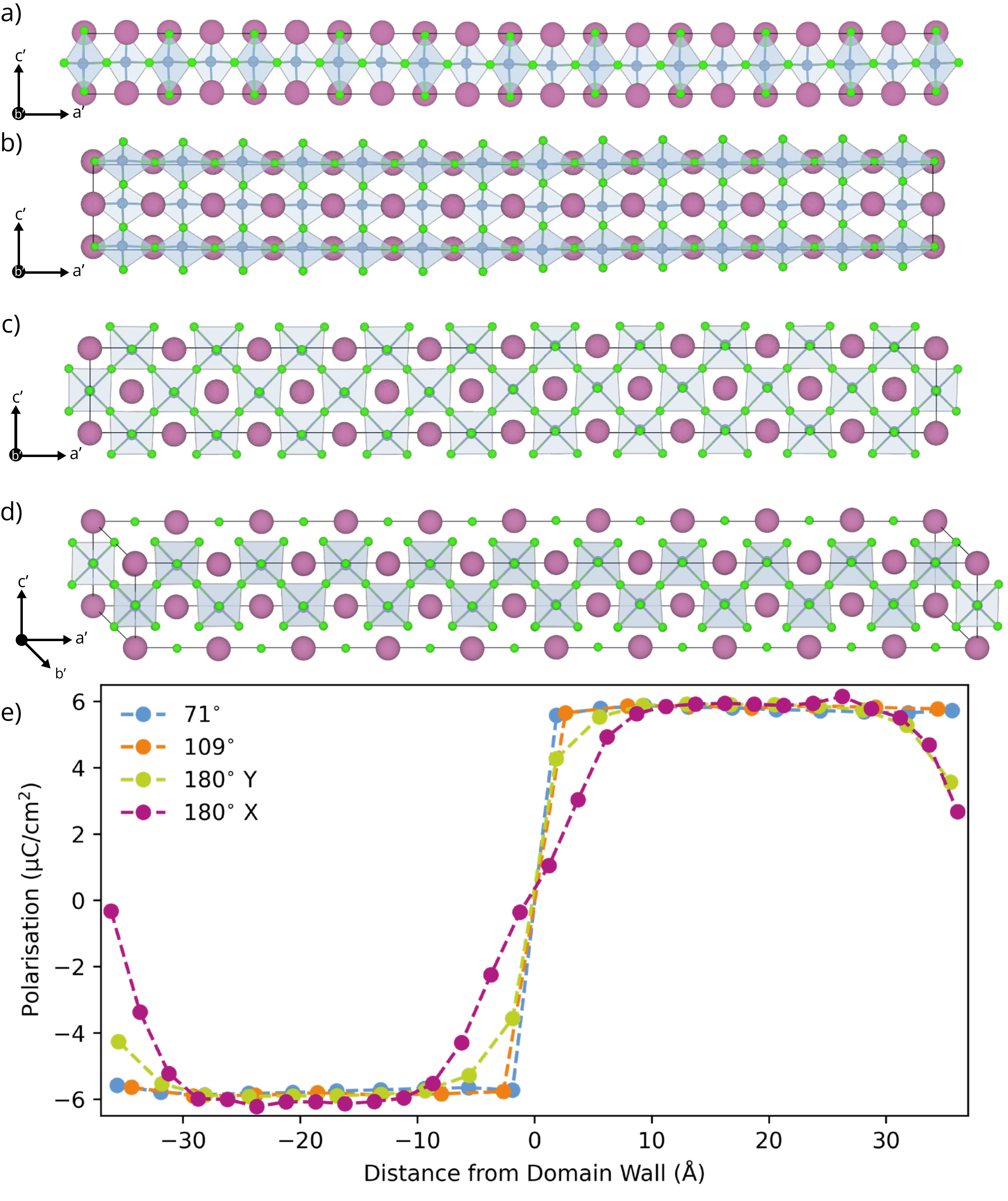}
    \caption{PBEsol optimised 71$\degree$ (a), 109$\degree$ (b), 180$\degree$ Y-type (c) and 180$\degree$ X-type (d) ferroelectric domain walls in \ce{CsGeCl3}. Estimated polarisations, from formal charges using the point charge model, are displayed in (e) as a function of the supercell dimension perpendicular to the domain walls. The relation between the supercells' and primitive cells lattice parameters is shown in SI Table S15.}
    \label{fig:CsGeCl3_DWs_Pol}
\end{figure}
 Calculated DW energies, summarised in Table \ref{tab:DW_energy}, indicate that 71$\degree$ DWs are the most favourable type in \ce{CsGeX3}, closely followed by 109$\degree$ DWs. Both exhibit minimal structural distortion and form narrow walls, as illustrated in SI Figures S12, S13, and S14 for \ce{CsGeCl3}, \ce{CsGeBr3}, and \ce{CsGeI3}, respectively. In contrast, 180$\degree$ DWs induce significant lattice distortions, resulting in substantially higher DW energies. Notably, the energies of 71$\degree$ and 109$\degree$ DWs in \ce{CsGeX3} are considerably lower than those reported for most ferroelectric oxide perovskites. For comparison, reported DW energies for 180$\degree$ DWs in \ce{LiNbO3}, 109$\degree$ DWs in \ce{BiFeO3}, 90$\degree$ DWs in \ce{PbTiO3}, 90$\degree$ DWs in orthorhombic \ce{KNbO3}, and 71$\degree$ and 109$\degree$ DWs in \ce{BaTiO3} are 141\cite{eggestad}, 33\cite{PhysRevLett.110.267601}, 35\cite{PhysRevB.65.104111}, 4.3\cite{PhysRevB.107.014101}, 3.8\cite{PhysRevB.86.155138} and 11.1\cite{PhysRevB.86.155138} mJ/m$^{2}$, respectively. In contrast, DW energies reported for the metal-organic halide perovskite \ce{MAPX3} (X = Cl, Br, I) are 3, 1, and 8 mJ/m$^2$ \cite{doi:10.1021/jz502666j}, respectively, which are comparable to those found in this study.
\begin{table}[htbp]
    \centering
    \caption{Calculated domain wall energies for DWs in \ce{CsGeCl3}, \ce{CsGeBr3} and \ce{CsGeI3} calculated using the PBEsol functional.}
    \begin{ruledtabular}
    \begin{tabular}{cccc}
    DW-type & \ce{CsGeCl3} (mJ/m$^2$) & \ce{CsGeBr3} (mJ/m$^2$) & \ce{CsGeI3} (mJ/m$^2$) \\
    \colrule
     71$\degree$ & 1.6 & 1.1 & 1.5 \\
     109$\degree$ & 1.7 & 4.2 & 2.5  \\
     180$\degree$ Y & 47 & 18 & 19 \\
     180$\degree$ X & 57 & 22 & 21 \\
    \end{tabular}
    \end{ruledtabular}
    \label{tab:DW_energy}
\end{table}

 As there are very little distortions associated with the low-energy DWs, there are almost no changes to the interatomic distances and bonding, and therefore almost no changes in the electronic structure are observed at the DWs, see SI Figure S15. Therefore, no reduction in bandgap at the DWs in \ce{CsGeX3} is expected and also not observed.
\subsection{Interactions Between Point Defects and Domain Walls}
 The 71$\degree$ DWs in \ce{CsGeCl3} were selected as the model system for investigating interactions between point defects and DWs, in order to reduce both complexity and computational cost. Given the similar behaviour of point defects (Figure \ref{fig:TL_Cl_I}, \ref{fig:defect_mig} and SI Figure S8) and DWs (Table \ref{tab:DW_energy}, SI Figure S12, S13 and S14) across these compounds, the interactions between point defects and DWs are also expected to be comparable. Moreover, as \ce{CsGeCl3} exhibits larger structural distortions, higher polarisation, stronger lone pair activity, and more pronounced polaronic behaviour, the interactions between point defects and DWs in \ce{CsGeCl3} is expected to be more significant than in the other compounds. The 71$\degree$ DW was chosen because it has the lowest formation energy for the compounds considered, making it the most likely to form. 

 SI Figure S16 shows normalised defect formation energies as a function of distance from a 71$\degree$ DW in \ce{CsGeCl3}. The energies are normalised to the defect formation of the same defect at the centre of one of the domains. None of the investigated point defects show any significant change in formation energy at DWs compared to domain centres or bulk. V$_{\text{Ge}}$ show the largest reduction in energy at the DW, but this is only a reduction of $\sim$0.005 eV, which is below the thermal energy ($k_BT$) at room temperature. Both V$_{\text{Cs}}$ and V$_{\text{Ge}}$ show a weak increase in energy approaching the DW before it decreases at the DW. However, these changes are very small and not likely to have any experimental significance.

 Point defects typically exhibit reduced defect formation energies at DWs, due to the inherent strain fields and distortions associated with the DWs \cite{Schaab2018,PhysRevResearch.2.033159,eggestad,10.1063/1.4789779}. Generally, point defects accumulate at DWs to maximise the volume of unperturbed bulk material, under the condition that the local structural distortions caused by DWs and the point defects are compatible \cite{PhysRevB.98.184102, PhysRevResearch.2.033159}. However, as there are very little distortions accompanying the low-energy DWs in \ce{CsGeX3}, almost no change in defect formation energies is observed at the 71$\degree$ DWs in \ce{CsGeCl3}.

\begin{figure}[ht]
    \centering
    \includegraphics[width=0.85\linewidth]{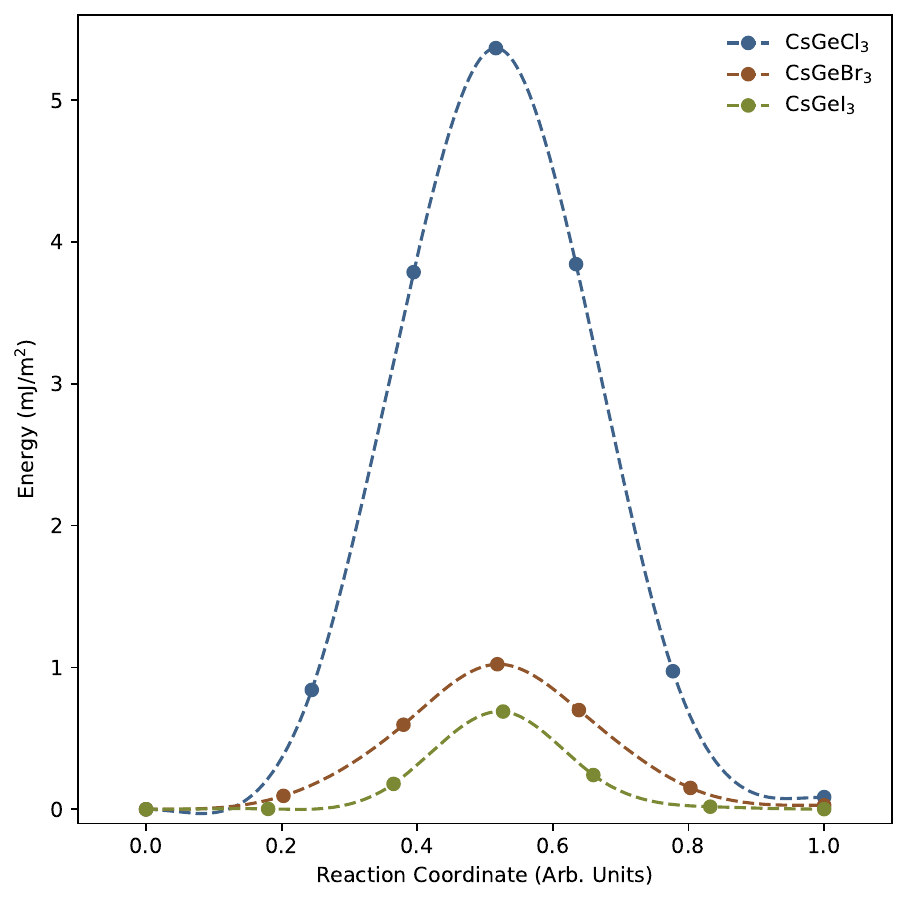}
    \caption{Calculated migration barriers for 71$\degree$ DWs in \ce{CsGeCl3}, \ce{CsGeBr3} and \ce{CsGeI3} using the PBEsol functional and the ci-NEB method.}
    \label{fig:DW_mig_pristine_71}
\end{figure}

\begin{figure*}
    \centering
    \includegraphics[width=0.95\textwidth]{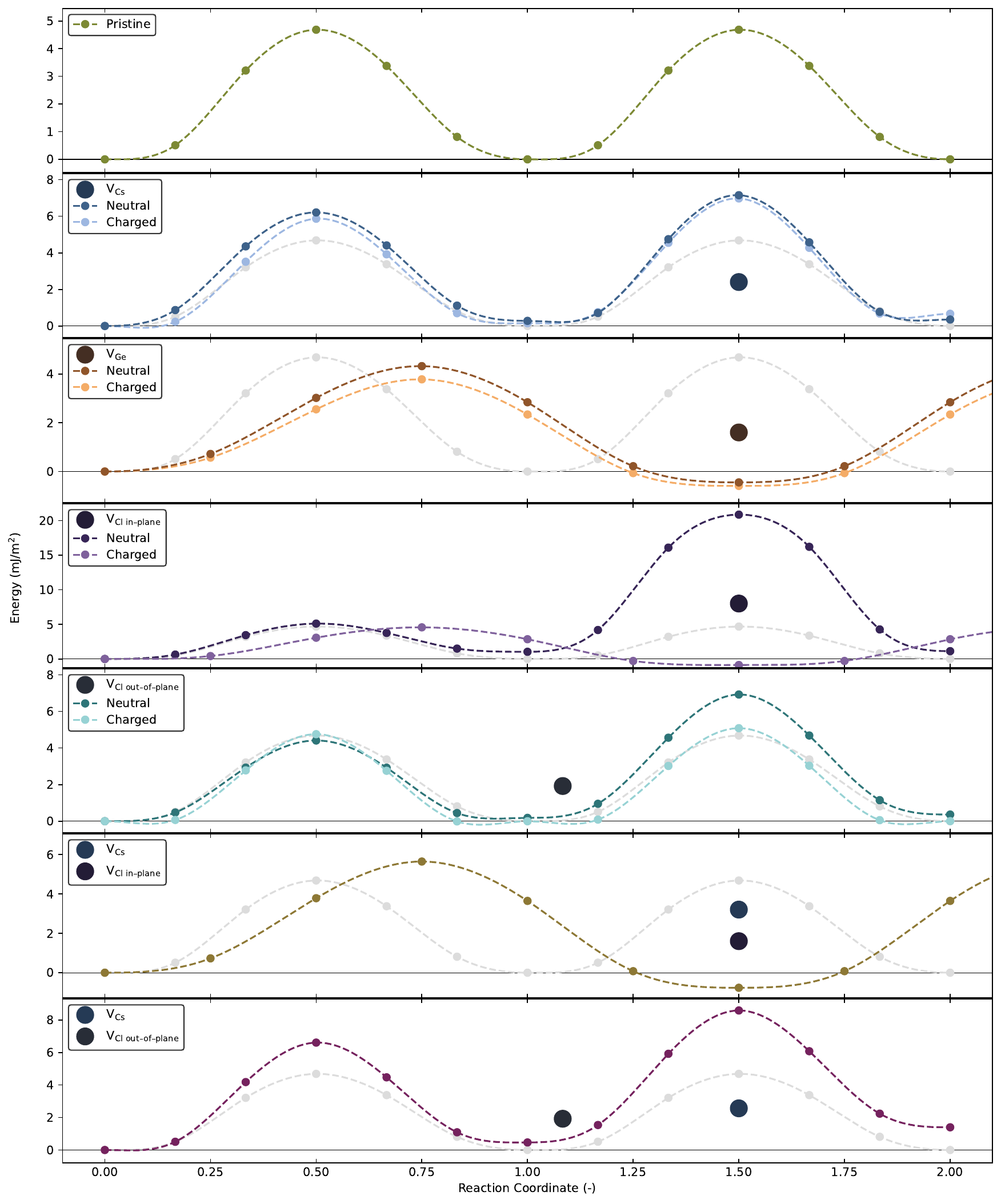}
    \caption{DW migration barriers for 71$\degree$ DWs in \ce{CsGeCl3} with and without defects in its vicinity. The first plot shows the migration barrier in the pristine DW cell. The next four plots show migration barriers in cells with V$_{\text{Cs}}$, V$_{\text{Ge}}$, V$_{\text{Cl}}$ and V$_{\text{Cl}}$, respectively. The gray dashed curves indicate the migration barrier in the pristine cell. For V$_{\text{Cs}}$, V$_{\text{Ge}}$, V$_{\text{Cl}}$, migration barriers are estimated both in neutral and charged cells. The two bottom panels display DW migration barriers with a defect pair in the cell. The filled circles indicate the position of the point defects.}
    \label{fig:dw_mobility}
\end{figure*}

 Migration barriers for 71$\degree$ DWs in \ce{CsGeCl3}, \ce{CsGeBr3} and \ce{CsGeI3} without point defects are displayed in Figure \ref{fig:DW_mig_pristine_71}. The figure shows very low migration barriers for 71$\degree$ DWs, with migration barriers of $\sim$ 5.4, 1.0 and 0.7 mJ/m$^2$ for \ce{CsGeCl3}, \ce{CsGeBr3} and \ce{CsGeI3}, respectively. These are all significantly lower than values reported for ferroelectric oxides, where, for example, 180$\degree$ DWs in \textit{h}-\ce{YMnO3} \ce{LiNbO3}, and \ce{PbTiO3} are reported with migration barriers of $\sim$30\cite{Kumagai2013,PhysRevResearch.2.033159}, 35\cite{eggestad} and 37\cite{PhysRevB.65.104111} mJ/m$^2$. \ce{CsGeCl3} show a migration barrier larger than the energy required to form the DW. Calculating the migration barriers of 109$\degree$ DWs was also attempted, but while "moving", each DW just transforms into two 71$\degree$ DWs. 
 The weak interaction between DWs and point defects is also reflected in the calculated DW migration barriers shown in Figure \ref{fig:dw_mobility}. The figure shows calculated migration barriers for a 71$\degree$ DW in \ce{CsGeCl3} with and without point defects in the supercell. The position of the defects is indicated by the filled circles and the DWs are moved in two steps. First approaching the defect, then moving past it. The interactions between the DW and defects are investigated using neutral and charged cells. The point defects studied are V$_{\text{Cs}}$, V$_{\text{Ge}}$ and in-plane and out-of-plane V$_{\text{Cl}}$ as well as V$_{\text{Cs}}$ + V$_{\text{Cl}}$ defect pairs.

 V$_{\text{Cs}}$ results in an increased migration barrier of almost 40~\%, however, the migration barriers are still very small. There is also a subtle increase in the migration barrier for the DW moving to the defect. Additionally, introducing a compensating hole into the system does not alter the results.

 Also, for V$_{\text{Ge}}$ there is no significant difference between calculations performed with or without hole compensation. V$_{\text{Ge}}$ does not result in any increase in migration barriers, but, as seen in the figure, the DW prefers to be situated where the defect is located. A small reduction in the total energy is also observed, similar to what is shown in SI Figure S16, when the DW coincides with the position of the V$_{\text{Ge}}$.

 The two inequivalent V$_{\text{Cl}}$ affect the DW very differently. An in-plane V$_{\text{Cl}}$ compensated by an electron leads to a substantial increase of the migration barrier of about 400\%. In contrast, when the in-plane V$_{\text{Cl}}$ is not compensated by an electron, the DW prefers to be co-located with the defect.  For the out-of-plane V$_{\text{Cl}}$, there is no change in the migration barrier if the defect is not charge-compensated, while charge compensation results in a similar increase in the migration barrier as observed for V$_{\text{Cs}}$.

 During the movement of a 71$^\circ$ domain wall, one in-plane Ge-Cl bond is formed while another bond is broken. If a charge-compensated in-plane V$_{\text{Cl}}$ is present at the domain wall, the polaron that compensates the defect must also transfer from one Ge atom to another, explaining the substantially increased migration barrier resulting from this specific point defect.

 The investigated defect clusters show combined characteristics of the individual defects in the charged unit cells. V$_{\text{Cs}}$ together with in-plane V$_{\text{Cl}}$ show increased migration barriers, similar to V$_{\text{Cs}}$, and a preference for the in-plane V$_{\text{Cl}}$. V$_{\text{Cs}}$ together with out-of-plane V$_{\text{Cl}}$ show increased migration barriers relatively similar to just a combination of the increased migration barriers of the individual defects.
\section{Discussion}
 The potential of \ce{CsGeCl3} as a transparent $p$-type conductor has previously been discussed in work by Huang et al. \cite{Huang_2014}, who highlighted its high hole mobility as well as favourable formation energy and transition level of $p$-type V$_{\text{Ge}}$. Our hybrid functional calculations support their findings regarding the high hole mobility and low formation energy of V$_{\text{Ge}}$. However, using the HSE06 functional, we find that V$_{\text{Ge}}$ exhibits a significantly deeper transition level than previously reported ($\sim0.5$ eV) \cite{Huang_2014}. In contrast, V$_{\text{Cs}}$ displays both a shallow transition level and low formation energy. For \ce{CsGeI3}, both V$_{\text{Cs}}$ and V$_{\text{Ge}}$ show low formation energies, shallow transition levels, and even lower hole effective masses. These results indicate that both materials are promising candidates for $p$-type conductivity. Additionally, at specific chemical environments the energy of n-type defects, namely V$_{\text{X}}$, is sufficiently high that the introduction of a good p-type dopant can pin the Fermi level within the valence band. Regarding transparent conductivity, \ce{CsGeCl3} show a band gap close to the energy of visible light, and has been reported experimentally with an even larger band gap \cite{doi:10.1021/ic970659e} as well as forming colourless crystals \cite{Christensen}.
 
 Calculated migration barriers for V$_{\text{Cl}}$ and 71$\degree$ DWs over V$_{\text{Cl}}$ in neutral cells show larger migration barriers than migration barriers calculated using charged cells. This means that the extra charge compensating electron in neutral cells affects both the mobility of halogen vacancies as well as the DW mobility. This is likely due to the charge-compensating electrons populating anti-bonding Ge\,$p$ states, destabilising the lone pair. In the charged cells without charge-compensating electrons, the lone pair is able to rotate more freely about its parent Ge ion and we hypothesise that this helps stabilising V$_{\text{Cl}}$ in the intermediate states, thus reducing the energy of the transition state. 

 Furthermore, both V$_{\text{Cl}}$ migration through path II and DW migration past an in-plane V$_{\text{Cl}}$ in neutral cells, require electron transfer between two Ge atoms (see SI Figure S10). In \ce{CsGeCl3}, the deep transition level of V$_{\text{Cl}}$ suggests an electron transfer through polaron hopping. The migration barrier associated with the polaron hopping will add to the total migration barrier, which explains the increased energy barriers of these migrations. 
 
 In contrast, in \ce{CsGeI3}, there is no difference in V$_{\text{I}}$ migration barriers for the different paths (SI Figure S8). V$_{\text{I}}$ shows a significantly more shallow defect level than V$_{\text{Cl}}$ and using PBEsol, the compensating electron is predominantly delocalised and thus no increase in migration barrier for path II is observed. Ultimately, calculations with charge compensation by electrons are likely not important as these materials show a strong $p$-type character and V$_{\text{Cl}}$ will instead be compensated by other defects rather than electrons.

 DWs in \ce{CsGeX3} show low formation energies, as well as low migration barriers. However, because of the subtlety of the structural distortions across the DWs, no significant change in point defect formation energy or electronic structure is observed at the DW. It is therefore likely difficult to obtain exotic electronic properties at the DWs of these materials. Despite the low defect formation energies, shallow transition levels, low defect migration barriers, and good carrier mobilities, this means that \ce{CsGeX3} are not promising candidates for DW-based electronics.

 Instead, low DW migration barriers indicate low energy loss and joule heating during polarisation reversal, and are thus useful for applications where high switching frequencies are needed, e.g. for FeRAMs, GHz transducers, and electro-optics. 

\section{Conclusion} 
Using hybrid functional DFT we have found that ferroelectric inorganic halides \ce{CsGeX3} show highly mobile electrons and holes. The point defects V$_{\text{Cs}}$ and V$_{\text{X}}$ show low formation energies and low and very low, respectively, migration energy barriers. This implies that point defects should be very important to the functional properties of these materials. Furthermore, low formation energy $p$-type defects show shallow defect levels and there is a potential for $p$-type doping of this ferroelectric material with an electronic band gap almost large enough to make it transparent to visible light.

We have shown that 71$\degree$ exhibit very low formation energies, closely followed by 109$\degree$ DWs. This is a consequence of the very subtle structural distortions found at these DWs compared to bulk, in contrast to the pronounced distortions often observed in oxide ferroelectrics. The absence of significant strain fields means that point defects show little tendency to accumulate at the DWs, and the electronic structure closely resembles that of the bulk. Consequently, \ce{CsGeX3} is unlikely to be suitable for DW-based electronics.

However, the low formation energy of the DWs, and especially the remarkably high mobility of 71$\degree$ DWs, make \ce{CsGeX3} materials promising candidates for applications as ferroelectrics that require high-frequency switching with low energy dissipation. The insensitivity to point defects and the inclination towards $p$-type doping emphasises the different properties and technological potential of ferroelectric halide perovskites compared to oxide perovskites.
\section{Acknowledgements}
Financial support for this project was provided from the Research Council of Norway through projects no. 302506, 301954, and 354614. Computational resources were provided through project no. NN9264K at Sigma 2 the National Infrastructure for High-Performance Computing and Data Storage in Norway.

\clearpage
\bibliographystyle{unsrt}
\bibliography{references}

\end{document}



\title{{\Large Supplementary Information}\\[-0.4 em]Defects and Domain Walls in Soft Ferroelectric \ce{CsGeX3} (X = Cl, Br, I)}

\author{Kristoffer Eggestad}
\affiliation{Department of Materials Science and Engineering,\\[-0.8 em] NTNU - Norwegian University of Science and Technology, NO-7491 Trondheim, Norway}

\author{Benjamin A. D. Williamson}
\affiliation{Department of Materials Science and Engineering,\\[-0.8 em] NTNU - Norwegian University of Science and Technology, NO-7491 Trondheim, Norway}

\author{Sverre M. Selbach}
\affiliation{Department of Materials Science and Engineering,\\[-0.8 em] NTNU - Norwegian University of Science and Technology, NO-7491 Trondheim, Norway}


\maketitle

\beginsupplement

\sloppy

\begin{figure}[ht]
    \centering
    \includegraphics[width=0.9\linewidth]{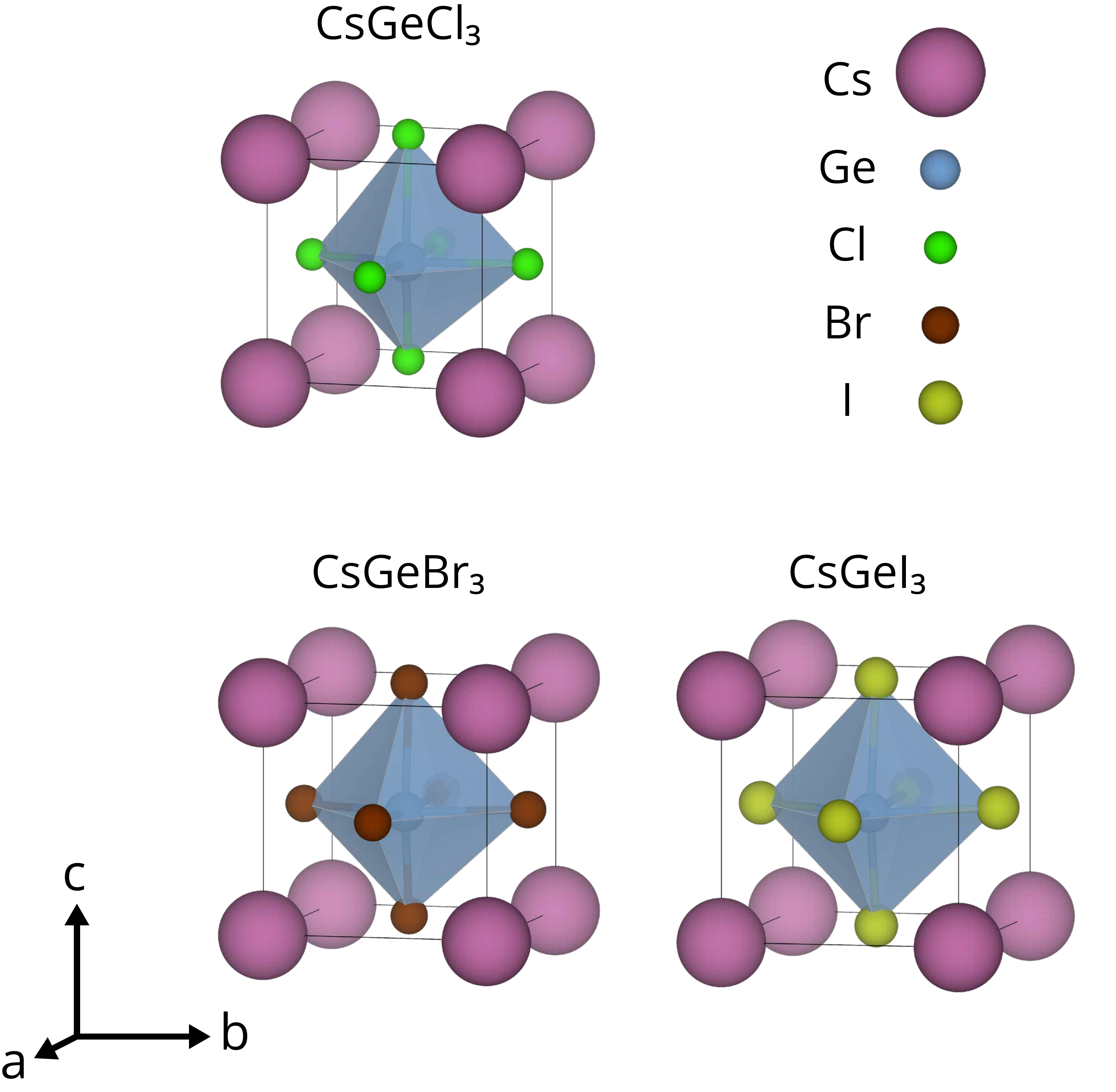}
    \caption{DFT optimised primitive structures for \ce{CsGeCl3}, \ce{CsGeBr3} and \ce{CsGeI3}.}
    \label{fig:prim_structs}
\end{figure}

\clearpage

\section{HSE06}

SI Figure \ref{fig:prim_structs} shows primitive unit cells of \ce{CsGeCl3}, \ce{CsGeBr3} and \ce{CsGeI3} optimised using hybrid DFT. Lattice parameters for the primitive unit cells calculated using different functionals are compared with experimental values in Table \ref{main-tab:lattice_comp}. In general, the HSE06 functional agrees best with experimental data and HSE06 optimised lattice parameters and atomic positions for the conventional unit cells are displayed in SI Table \ref{tab:lattice_para}, \ref{tab:atomic_pos_Cl}, \ref{tab:atomic_pos_Br} and \ref{tab:atomic_pos_I}. 

\begin{table}[ht]
    \centering

    \caption{DFT optimised lattice parameters using HSE06 for the conventional structures of \ce{CsGeCl3}, \ce{CsGeBr3} and \ce{CsGeI3}.}
    \begin{tabular}{cccc}
        \toprule
        & \ce{CsGeCl3} & \ce{CsGeBr3} & \ce{CsGeI3} \\
        \midrule
       a, b  & 7.761096 Å & 7.998540 Å & 8.459225 Å \\
       c & 9.632980 Å & 10.183176 Å & 10.869453 Å \\
       $\alpha$, $\beta$ & 90$\degree$ & 90$\degree$ & 90$\degree$ \\
       $\gamma$ & 120$\degree$ & 120$\degree$ & 120$\degree$ \\
        \bottomrule
    \end{tabular}
    \label{tab:lattice_para}
\end{table}

\begin{table}[ht]
    \centering

    \caption{DFT optimised atomic positions using HSE06 for the conventional \ce{CsGeCl3} structure.}
    \begin{tabular}{ccccc}\toprule
       Specie  & a & b & c & Wyckoff \\
       \midrule
       Cs  & 1/3 & 2/3 & 0.655318 & 3a \\
       Ge  & 1/3 & 2/3 & 0.139602   & 3a \\
       Cl  & 0.482434   &  0.964867 & 0.008805 & 9b \\
        \bottomrule
    \end{tabular}
    \label{tab:atomic_pos_Cl}
\end{table}

\begin{table}[ht]
    \centering

    \caption{DFT optimised atomic positions using HSE06 for the conventional \ce{CsGeBr3} structure.}
    \begin{tabular}{ccccc}\toprule
       Specie  & a & b & c & Wyckoff \\
       \midrule
       Cs  & 1/3 & 2/3 & 0.662909   & 3a \\
       Ge  & 1/3 & 2/3 & 0.13601   & 3a \\
       Br  & 0.490037   &  0.980074   & 0.007472 & 9b \\
        \bottomrule
    \end{tabular}
    \label{tab:atomic_pos_Br}
\end{table}

\begin{table}[ht]
    \centering

    \caption{DFT optimised atomic positions using HSE06 for the conventional \ce{CsGeI3} structure.}
    \begin{tabular}{ccccc}\toprule
       Specie  & a & b & c & Wyckoff \\
       \midrule
       Cs  & 1/3 & 2/3 & 0.666054   & 3a \\
       Ge  & 1/3 & 2/3 & 0.133573   & 3a \\
       I  & 0.495887   &  0.991775 & 0.007236 & 9b \\
        \bottomrule
    \end{tabular}
    \label{tab:atomic_pos_I}
\end{table}

Calculated Bader charges, displayed in SI Table \ref{tab:bader_HSE06}, show a decrease in the ionic nature of \ce{CsGeX3} from the chloride to the iodide. Similar results, achieved using the PBEsol functional, are displayed in SI Table \ref{tab:bader_PBEsol}.

\begin{table}[htbp]
    \centering
    \caption{Bader charges calculated using the HSE06 functional.}
    \begin{tabular}{cccc}
        \toprule
         & \ce{CsGeCl3} & \ce{CsGeBr3} & \ce{CsGeI3} \\
         \midrule
        Cs & 0.91 & 0.89 & 0.87 \\
        Ge & 1.19 & 0.94 & 0.72 \\
        X & -0.70 & -0.61 & -0.53 \\
        \bottomrule
    \end{tabular}

    \label{tab:bader_HSE06}
\end{table}

SI Table \ref{tab:BEC_all_HSE06}, \ref{tab:diel_HSE06}, \ref{tab:piezo_HSE06} and \ref{tab:elastic_HSE06} show Born effective charges (BEC), dielectric constants, piezoelectric and elastic tensors from finite differences calculations using the HSE06 functional. The values are displayed for conventional unit cells.

\begin{table}[ht]
    \centering

    \caption{Born Effective charges (BEC) calculated using finite differences with the HSE06 functional. The BEC are displayed for the conventional unit cells.}
    \begin{tabular}{cccc}
    \toprule
    Specie & \ce{CsGeCl3} & \ce{CsGeBr3} & \ce{CsGeI3} \\
    \midrule
     Cs    & 
$
\begin{pmatrix}
    1.38  & 0 & \\
    0  &  1.38 & 0 \\
    0 & 0 &  1.36 \\
\end{pmatrix}
$
        &
        $
\begin{pmatrix}
    1.44  & 0 & \\
    0  &  1.44 & 0 \\
    0 & 0 &  1.40 \\
\end{pmatrix}
$
&
$
\begin{pmatrix}
    1.46  & 0 & \\
    0  &  1.46 & 0 \\
    0 & 0 &  1.41 \\
\end{pmatrix}
$\\
     Ge    & 
$
\begin{pmatrix}
    2.97  & 0 & \\
    0  &  2.97 & 0 \\
    0 & 0 &  2.52 \\
\end{pmatrix}
$
        &
        $
\begin{pmatrix}
    3.36  & 0 & \\
    0  &  3.36 & 0 \\
    0 & 0 &  2.76 \\
\end{pmatrix}
$
&
$
\begin{pmatrix}
    3.81  & 0 & \\
    0  &  3.81 & 0 \\
    0 & 0 &  2.89 \\
\end{pmatrix}
$\\
     X    & 
$
\begin{pmatrix}
    -1.45  & 0 & \\
    0  &  -1.45 & 0 \\
    0 & 0 &  -1.29 \\
\end{pmatrix}
$
        &
        $
\begin{pmatrix}
    -1.60  & 0 & \\
    0  &  -1.60 & 0 \\
    0 & 0 &  -1.38 \\
\end{pmatrix}
$
&
$
\begin{pmatrix}
    -1.76  & 0 & \\
    0  &  -1.76 & 0 \\
    0 & 0 &  -1.43 \\
\end{pmatrix}
$\\
\bottomrule
    \end{tabular}
    \label{tab:BEC_all_HSE06}
\end{table}

\begin{table}[ht]
    \centering

    \caption{Dielectric constants for the conventional unit cells of \ce{CsGeCl3},\ce{CsGeBr3} and \ce{CsGeI3} calculated using finite differences with the HSE06 functional.}
    \begin{tabular}{ccc}
    \toprule
     \ce{CsGeCl3} & \ce{CsGeBr3} & \ce{CsGeI3} \\
     \midrule
$
\begin{pmatrix}
    13.63  & 0 & \\
    0  &  13.63 & 0 \\
        0 & 0 &  11.82 \\
\end{pmatrix}
$
        &
        $
\begin{pmatrix}
13.85 & 0 & 0 & \\
0 & 13.85 & 0 & \\
0 & 0 & 14.86 & \\
\end{pmatrix}
$
&
$
\begin{pmatrix}
14.88 & 0 & 0 & \\
0 & 14.88 & 0 & \\
0 & 0 & 19.54 & \\
\end{pmatrix}
$\\     
     \bottomrule
    \end{tabular}
    
    \label{tab:diel_HSE06}
\end{table}

\begin{table}[ht]
    \centering
    \caption{Piezoelectric tensors for the conventional unit cell calculated using finite differences with the HSE06 functional.}
    \begin{tabular}{cc}
    \toprule
       Material  & Piezoelectric Tensor (Voigt) (C/m$^2$) \\
    \midrule
\ce{CsGeCl3}
         & 
$
\begin{pmatrix}
0 & 0 & 0 & 0 & 0.32 & 0.25 \\
0.25 & -0.25 & 0 & 0.32 & 0 & 0 \\
0.25 & 0.25 & 0.48 & 0 & 0 & 0 \\
\end{pmatrix}
$
\\
\ce{CsGeBr3}
         & 
$
\begin{pmatrix}
0 & 0 & 0 & 0 & 0.41 & 0.28 \\
0.28 & -0.28 & 0 & 0.41 & 0 & 0 \\
0.27 & 0.27 & 0.5 & 0 & 0 & 0 \\
\end{pmatrix}
$
\\
\ce{CsGeI3}
         & 
$
\begin{pmatrix}
0 & 0 & 0 & 0 & 0.45 & 0.32 \\
0.32 & -0.32 & 0 & 0.45 & 0 & 0 \\
0.29 & 0.29 & 0.56 & 0 & 0 & 0 \\
\end{pmatrix}
$
\\
    \bottomrule
    \end{tabular}
    
    \label{tab:piezo_HSE06}
\end{table}

\begin{table}[ht]
    \centering
    \caption{Elastic tensors for the conventional unit cell calculated using finite differences with the HSE06 functional.}
    \begin{tabular}{cc}
    \toprule
       Material  & Elastic Tensor (Voigt) (GPa) \\
    \midrule
\ce{CsGeCl3}
         & 
$
\begin{pmatrix}
22.24 & 12.26 & 9.79 & 1.57 & 0 & 0 \\
12.26 & 22.24 & 9.79 & -1.57 & 0 & 0 \\
9.79 & 9.79 & 16.06 & 0 & 0 & 0 \\
1.57 & -1.57 & 0 & 7.21 & 0 & 0 \\
0 & 0 & 0 & 0 & 7.21 & 1.57 \\
0 & 0 & 0 & 0 & 1.57 & 4.99 \\
\end{pmatrix}
$
\\
\ce{CsGeBr3}
         & 
$
\begin{pmatrix}
18.38 & 10.14 & 9.16 & 2.02 & 0 & 0 \\
10.14 & 18.38 & 9.16 & -2.02 & 0 & 0 \\
9.16 & 9.16 & 13.13 & 0 & 0 & 0 \\
2.02 & -2.02 & 0 & 6.33 & 0 & 0 \\
0 & 0 & 0 & 0 & 6.33 & 2.02 \\
0 & 0 & 0 & 0 & 2.02 & 4.12 \\
\end{pmatrix}
$
\\
\ce{CsGeI3}
         & 
$
\begin{pmatrix}
13.37 & 6.44 & 6.79 & 2.62 & 0 & 0 \\
6.44 & 13.37 & 6.79 & -2.62 & 0 & 0 \\
6.79 & 6.79 & 9.12 & 0 & 0 & 0 \\
2.62 & -2.62 & 0 & 5.71 & 0 & 0 \\
0 & 0 & 0 & 0 & 5.71 & 2.62 \\
0 & 0 & 0 & 0 & 2.62 & 3.46 \\
\end{pmatrix}
$
\\
    \bottomrule
    \end{tabular}
    
    \label{tab:elastic_HSE06}
\end{table}


\clearpage
\section{PBEsol}
\label{sec:PBEsol}

SI Table \ref{tab:BEC_all}, \ref{tab:diel_PBEsol}, \ref{tab:piezo_PBEsol} and \ref{tab:elastic_PBEsol} display born BEC, dielectric constants, piezoelectric and elastic tensors for the conventional unit cells of \ce{CsGeCl3},\ce{CsGeBr3} and \ce{CsGeI3}. BEC, piezoelectric and elastic tensors, as well as the ionic contribution to the dielectric constants, are calculated using density functional perturbation theory (DFPT) with the PBEsol functional.

\begin{table}[ht]
    \centering

    \caption{Bader charges calculated using the PBEsol functional.}
    \begin{tabular}{cccc}
        \toprule
         Specie & \ce{CsGeCl3} & \ce{CsGeBr3} & \ce{CsGeI3} \\
         \midrule
        Cs & 0.87 & 0.85 & 0.82 \\
        Ge & 1.07 & 0.94 & 0.64 \\
        X & -0.65 & -0.60 & -0.49 \\
        \bottomrule
    \end{tabular}

    \label{tab:bader_PBEsol}
\end{table}

\begin{table}[ht]
    \centering

    \caption{Born Effective charges (BEC) calculated using DFPT with the PBEsol functional. The BEC are displayed for the conventional unit cells.}
    \begin{tabular}{cccc}
    \toprule
    Specie & \ce{CsGeCl3} & \ce{CsGeBr3} & \ce{CsGeI3} \\
    \midrule
     Cs    & 
$
\begin{pmatrix}
    1.36  & 0 & \\
    0  &  1.36 & 0 \\
    0 & 0 &  1.36 \\
\end{pmatrix}
$
        &
        $
\begin{pmatrix}
    1.38  & 0 & \\
    0  &  1.38 & 0 \\
    0 & 0 &  1.37 \\
\end{pmatrix}
$
&
$
\begin{pmatrix}
    1.43  & 0 & \\
    0  &  1.43 & 0 \\
    0 & 0 &  1.42 \\
\end{pmatrix}
$\\
     Ge    & 
$
\begin{pmatrix}
    3.80  & 0 & \\
    0  &  3.80 & 0 \\
    0 & 0 &  3.56 \\
\end{pmatrix}
$
        &
        $
\begin{pmatrix}
    4.69  & 0 & \\
    0  &  4.69 & 0 \\
    0 & 0 &  4.40 \\
\end{pmatrix}
$
&
$
\begin{pmatrix}
    5.43  & 0 & \\
    0  &  5.43 & 0 \\
    0 & 0 &  4.81 \\
\end{pmatrix}
$\\
     X    & 
$
\begin{pmatrix}
    -1.72  & 0 & \\
    0  &  -1.72 & 0 \\
    0 & 0 &  -1.64 \\
\end{pmatrix}
$
        &
        $
\begin{pmatrix}
    -2.02  & 0 & \\
    0  &  -2.02 & 0 \\
    0 & 0 &  -1.92 \\
\end{pmatrix}
$
&
$
\begin{pmatrix}
    -2.29  & 0 & \\
    0  &  -2.29 & 0 \\
    0 & 0 &  -2.08 \\
\end{pmatrix}
$\\
\bottomrule
    \end{tabular}
    \label{tab:BEC_all}
\end{table}

\begin{table}[ht]
    \centering

    \caption{Dielectric constants for the conventional unit cells of \ce{CsGeCl3},\ce{CsGeBr3} and \ce{CsGeI3} calculated using DFPT with the PBEsol functional.}
    \begin{tabular}{ccc}
    \toprule
     \ce{CsGeCl3} & \ce{CsGeBr3} & \ce{CsGeI3} \\
     \midrule
$
\begin{pmatrix}
16.97 & 0 & 0 \\
0 & 16.97 & 0 \\
0 & 0 & 12.9 \\
\end{pmatrix}
$
        &
        $
\begin{pmatrix}
38.07 & 0 & 0 & \\
0 & 38.07 & 0 & \\
0 & 0 & 28.86 & \\
\end{pmatrix}
$
&
$
\begin{pmatrix}
40.19 & 0 & 0 & \\
0 & 40.19 & 0 & \\
0 & 0 & 30.78 & \\
\end{pmatrix}
$\\     
     \bottomrule
    \end{tabular}
    
    \label{tab:diel_PBEsol}
\end{table}

\begin{table}[ht]
    \centering
    \caption{Piezoelectric tensors for the conventional unit cell calculated using DFPT with the PBEsol functional.}
    \begin{tabular}{cc}
    \toprule
       Material  & Piezoelectric Tensor (Voigt) (C/m$^2$) \\
    \midrule
\ce{CsGeCl3}
         & 
$
\begin{pmatrix}
0 & 0 & 0 & 0 & 0.68 & 0.47 \\
0.47 & -0.47 & 0 & 0.68 & 0 & 0 \\
0.77 & 0.77 & 0.86 & 0 & 0 & 0 \\
\end{pmatrix}
$
\\
\ce{CsGeBr3}
         & 
$
\begin{pmatrix}
0 & 0 & 0 & 0 & 1.03 & 0.76 \\
0.76 & -0.76 & 0 & 1.03 & 0 & 0 \\
1.11 & 1.11 & 1.44 & 0 & 0 & 0 \\
\end{pmatrix}
$
\\
\ce{CsGeI3}
         & 
$
\begin{pmatrix}
0 & 0 & 0 & 0 & 1.02 & 0.78 \\
0.78 & -0.78 & 0 & 1.02 & 0 & 0 \\
0.63 & 0.63 & 0.99 & 0 & 0 & 0 \\
\end{pmatrix}
$
\\
    \bottomrule
    \end{tabular}
    
    \label{tab:piezo_PBEsol}
\end{table}

\begin{table}[ht]
    \centering
    \caption{Elastic tensors for the conventional unit cell calculated using DFPT with the PBEsol functional.}
    \begin{tabular}{cc}
    \toprule
       Material  & Elastic Tensor (Voigt) (GPa) \\
    \midrule
\ce{CsGeCl3}
         & 
$
\begin{pmatrix}
20.54 & 12.74 & 12.12 & 2.4 & 0 & 0 \\
12.74 & 20.54 & 12.12 & -2.4 & 0 & 0 \\
12.12 & 12.12 & 16.89 & 0 & 0 & 0 \\
2.4 & -2.4 & 0 & 6.51 & 0 & 0 \\
0 & 0 & 0 & 0 & 6.51 & 2.4 \\
0 & 0 & 0 & 0 & 2.4 & 3.9 \\
\end{pmatrix}
$
\\
\ce{CsGeBr3}
         & 
$
\begin{pmatrix}
12.82 & 6.47 & 9.93 & 2.94 & 0 & 0 \\
6.47 & 12.82 & 9.93 & -2.94 & 0 & 0 \\
9.93 & 9.93 & 14.86 & 0 & 0 & 0 \\
2.94 & -2.94 & 0 & 5.26 & 0 & 0 \\
0 & 0 & 0 & 0 & 5.26 & 2.94 \\
0 & 0 & 0 & 0 & 2.94 & 3.18 \\
\end{pmatrix}
$
\\
\ce{CsGeI3}
         & 
$
\begin{pmatrix}
8.06 & 1.04 & 5.05 & 3.49 & 0 & 0 \\
1.04 & 8.06 & 5.05 & -3.49 & 0 & 0 \\
5.05 & 5.05 & 8.64 & 0 & 0 & 0 \\
3.49 & -3.49 & 0 & 5.35 & 0 & 0 \\
0 & 0 & 0 & 0 & 5.35 & 3.49 \\
0 & 0 & 0 & 0 & 3.49 & 3.51 \\
\end{pmatrix}
$
\\
    \bottomrule
    \end{tabular}
    
    \label{tab:elastic_PBEsol}
\end{table}

\clearpage

\section{Electronic Structure}

\begin{figure}[ht]
    \centering

    \includegraphics[width=\columnwidth]{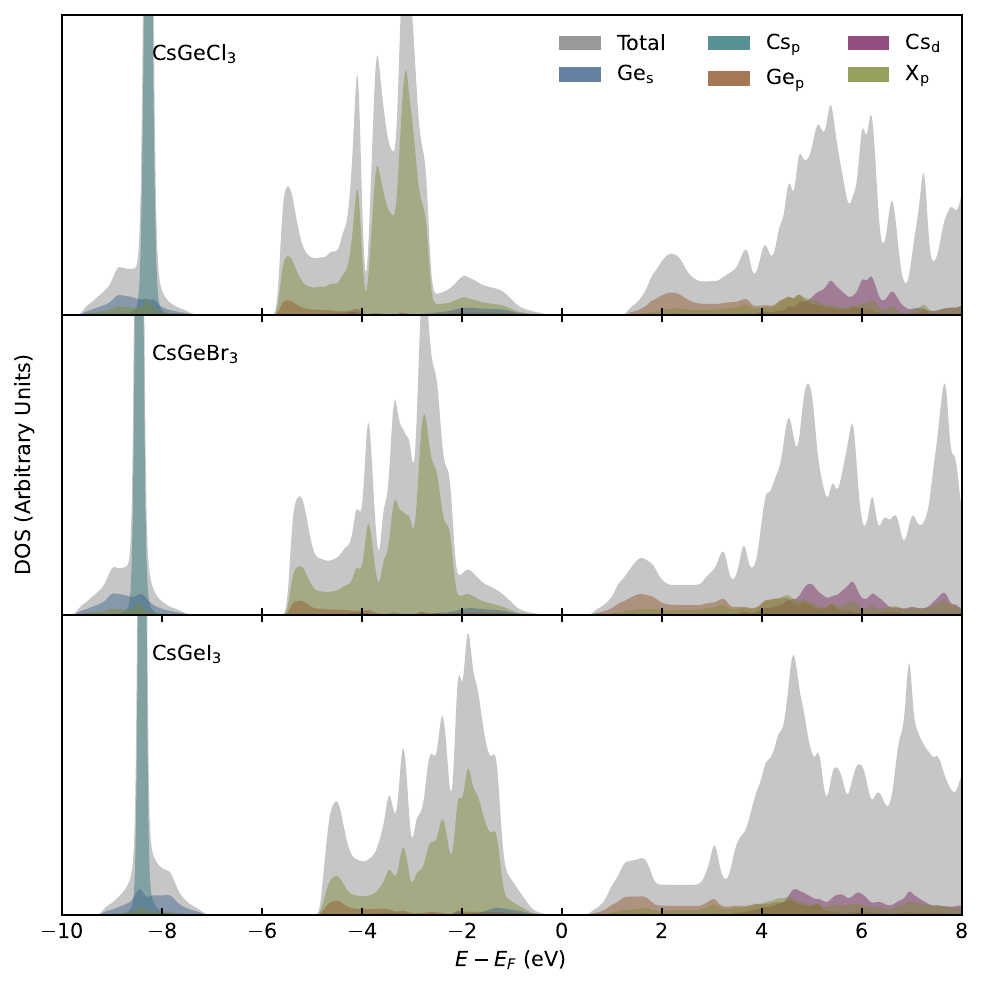}
    \caption{Electronic DOS calculated using the PBEsol functional.}
    \label{fig:DOS_all_PBEsol}
\end{figure}

\begin{figure}[ht]
    \centering
    \includegraphics[width=\columnwidth]{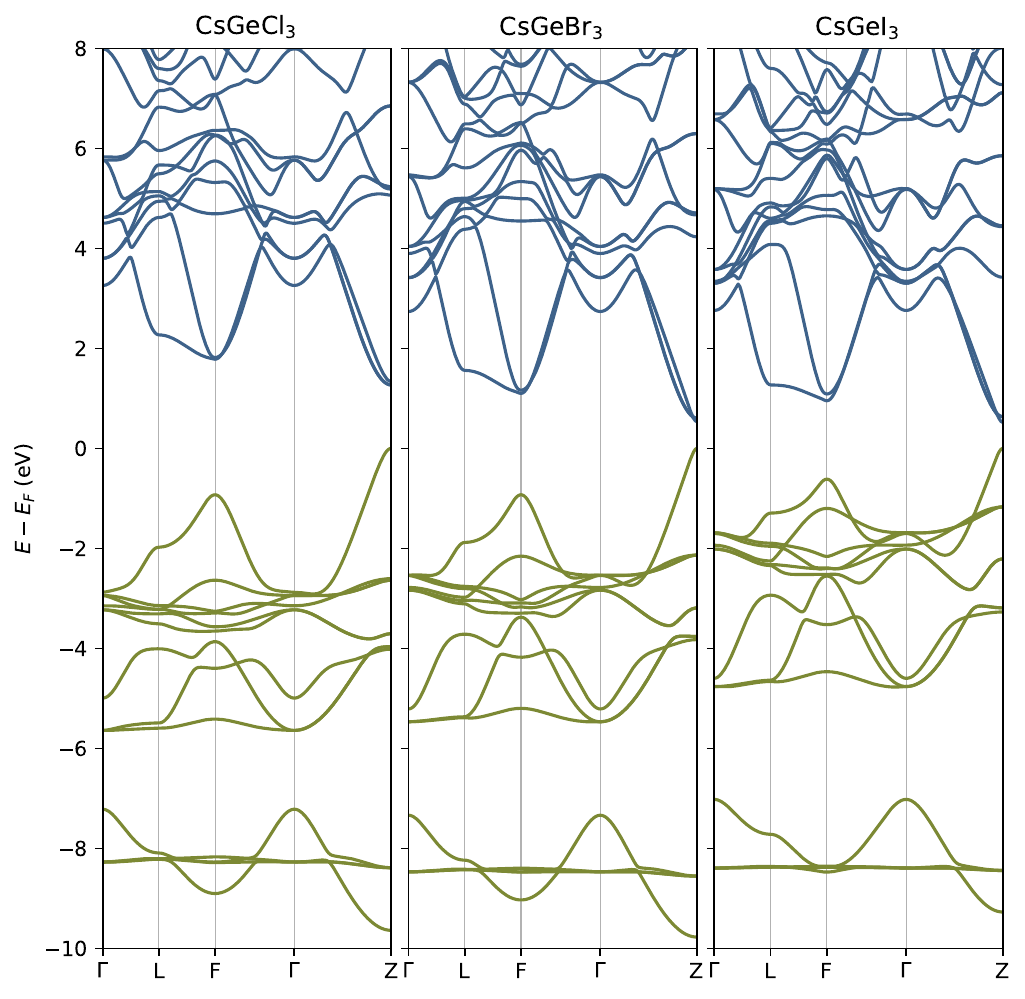}
    \caption{Electronic band structure calculated using the PBEsol functional.}
    \label{fig:band_all_PBEsol}
\end{figure}

\begin{figure}[ht]
    \centering

    \includegraphics[width=\columnwidth]{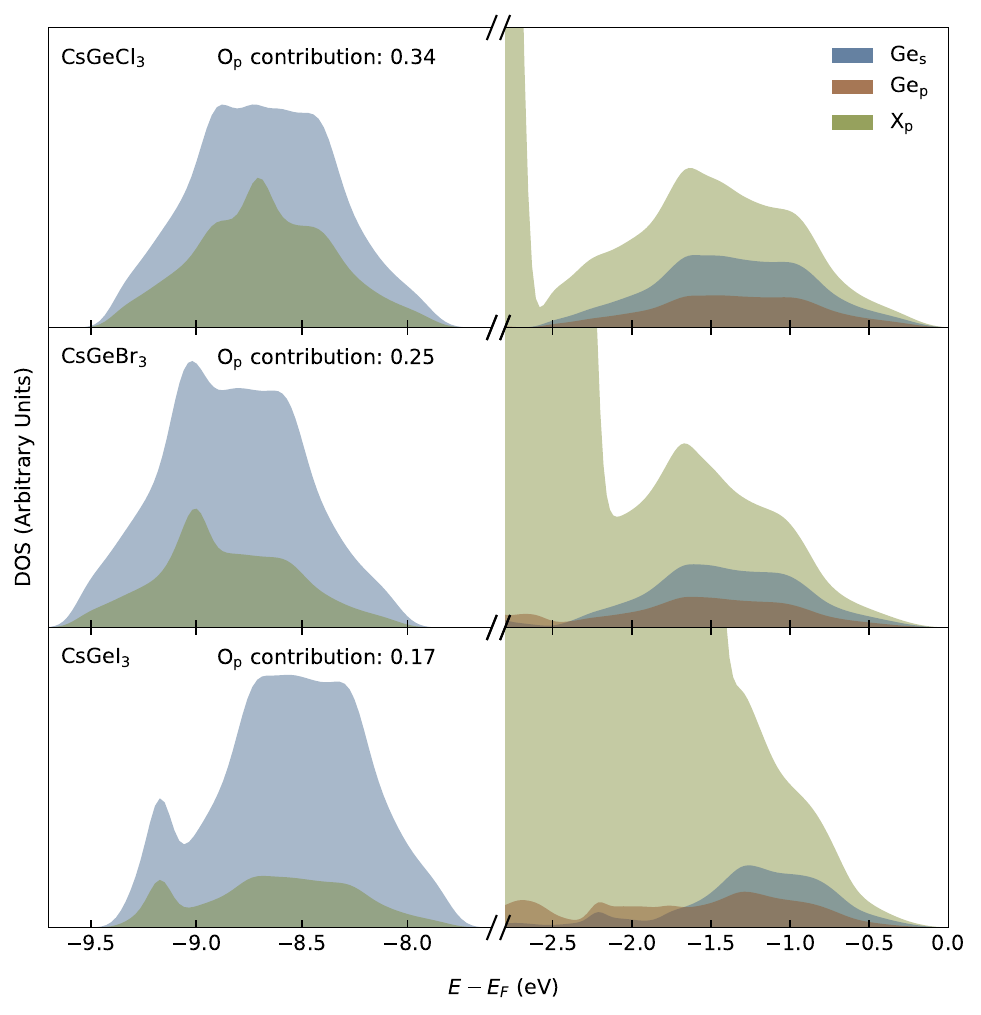}
    \caption{Electronic DOS calculated using the HSE06 functional showing the lone pair interaction.}
    \label{fig:DOS_all_zoomed}
\end{figure}

SI Figure \ref{fig:DOS_all_PBEsol} and \ref{fig:band_all_PBEsol} show electronic DOSes and band structures calculated using the PBEsol functional. The electronic structures are similar to those calculated using the HSE06 functional (main text Figure \ref{main-fig:dos_all} and \ref{main-fig:band_all}), albeit show significantly smaller band gaps. SI Figure \ref{fig:DOS_all_zoomed} displays portions of the HSE06 calculated DOSes showing the lone pair interactions together with the quantified degree of mixing of \ce{Ge}$_{\text{s}}$ and \ce{X}$_{\text{p}}$ following work by Payne et al.\cite{PhysRevLett.96.157403}.

\begin{figure}[ht]
    \centering
    \includegraphics[width=0.9\linewidth]{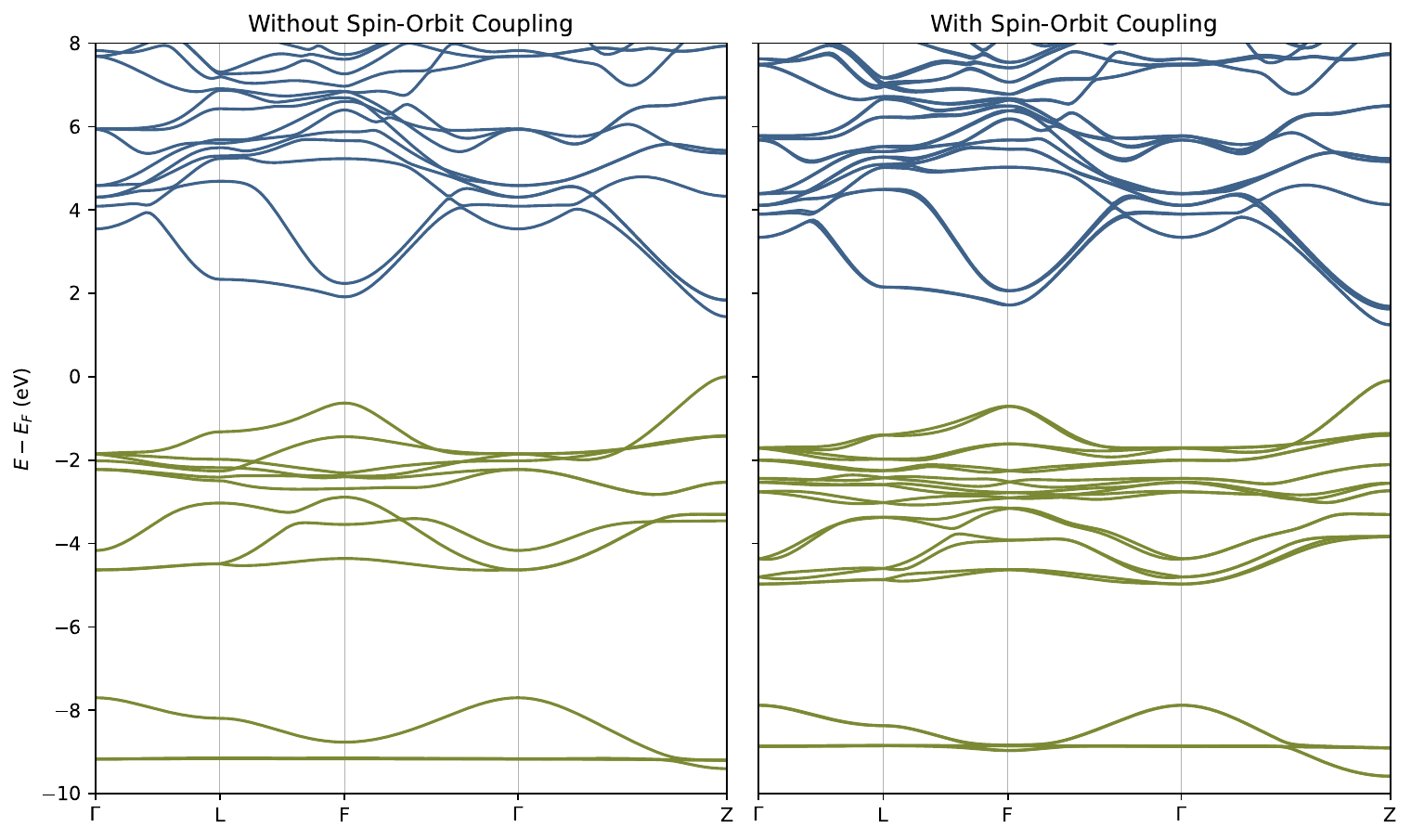}
    \caption{The band structure of \ce{CsGeI3} calculated using the HSE06 functional with and without spin-orbit coupling.}
    \label{fig:SOC_CsGeI3_band}
\end{figure}

\clearpage
\section{Defects}

\begin{figure}[ht]
    \centering
    \includegraphics[width=0.85\linewidth]{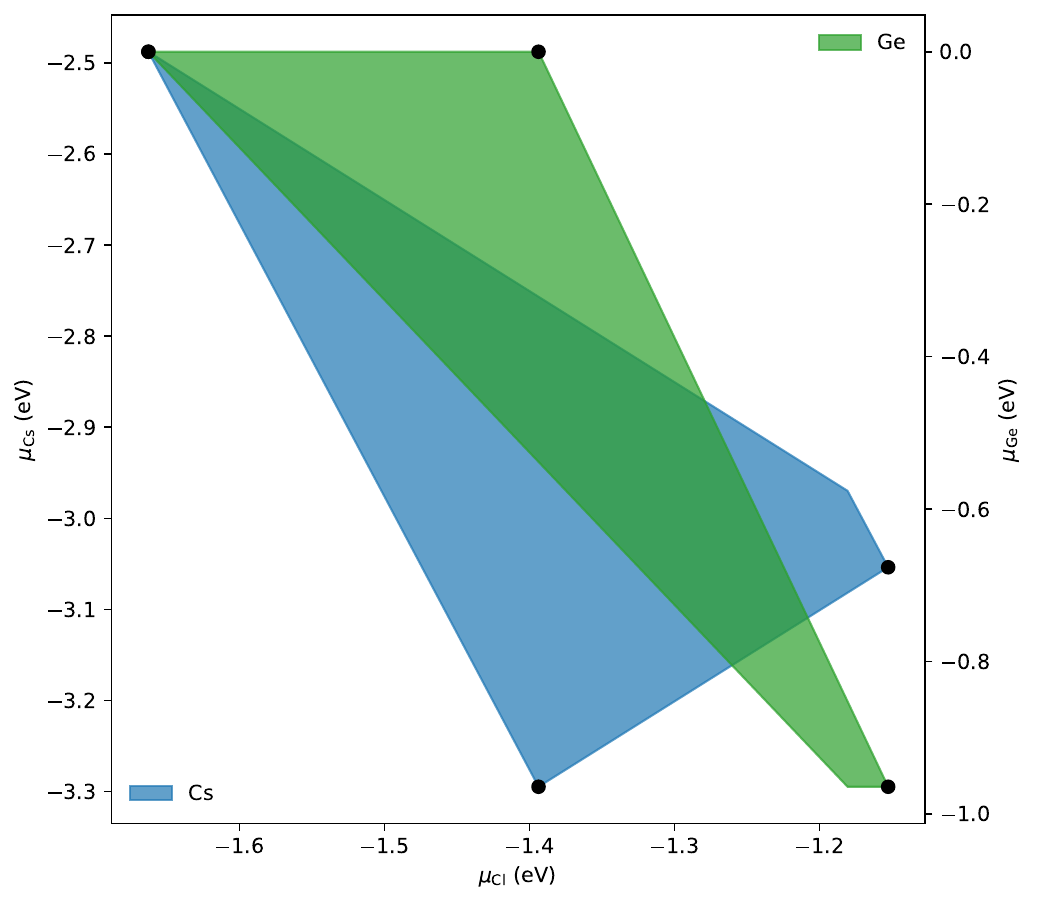} 
    \caption{Thermodynamic stability region calculated using the HSE06 functional for \ce{CsGeCl3}. The black dots indicate the chemical potentials used in Figure \ref{main-fig:TL_Cl_I} (a).}
    \label{fig:stab_CsGeCl3}
\end{figure}

\begin{figure}[ht]
    \centering
    \includegraphics[width=0.85\linewidth]{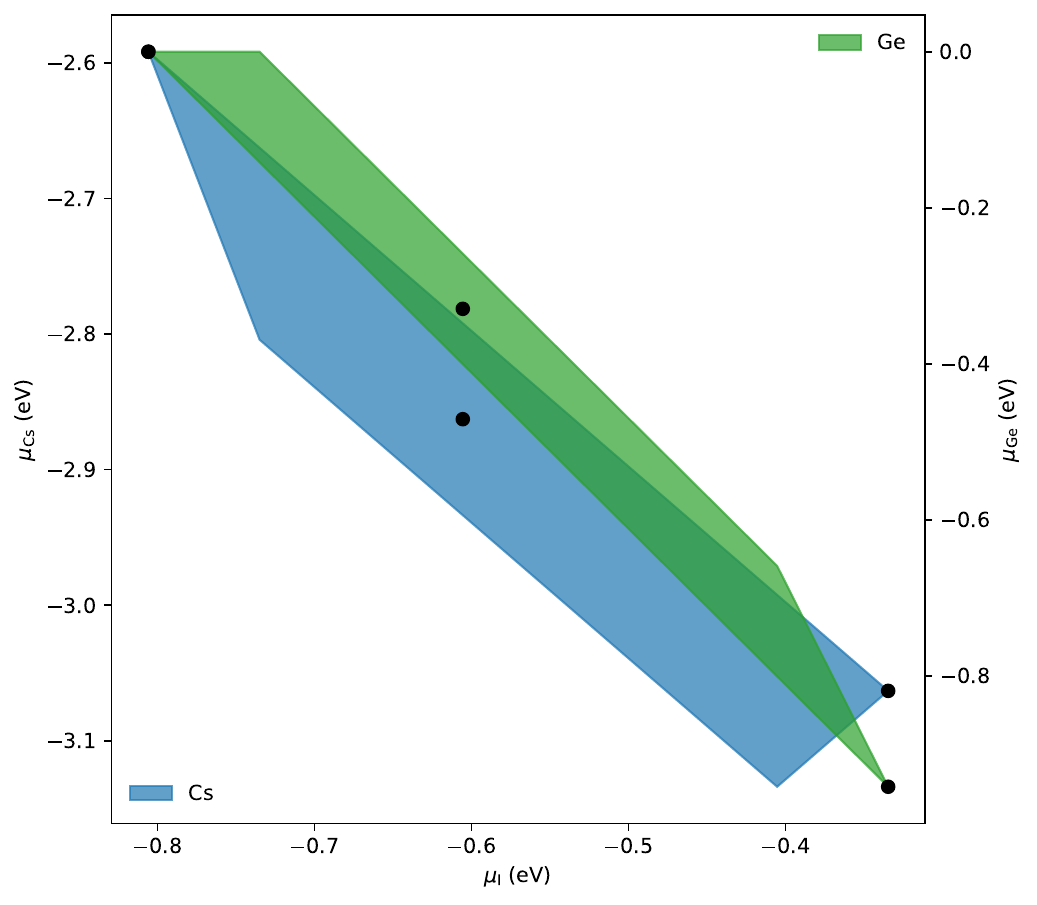} 
    \caption{Thermodynamic stability region calculated using the HSE06 functional for \ce{CsGeI3}. The black dots indicate the chemical potentials used in Figure \ref{main-fig:TL_Cl_I} (b).}
    \label{fig:stab_CsGeI3}
\end{figure}

Thermodynamic stability regions for \ce{CsGeCl3} and \ce{CsGeI3} are shown in SI Figure \ref{fig:stab_CsGeCl3} and \ref{fig:stab_CsGeI3}. These are calculated by evaluating total energies of all competing phases using the Chemical Potential Limits Analysis Program (\texttt{CPLAP}) \cite{cplap}.

\clearpage

\begin{figure}[ht]
    \centering
    \includegraphics[width=0.75\linewidth]{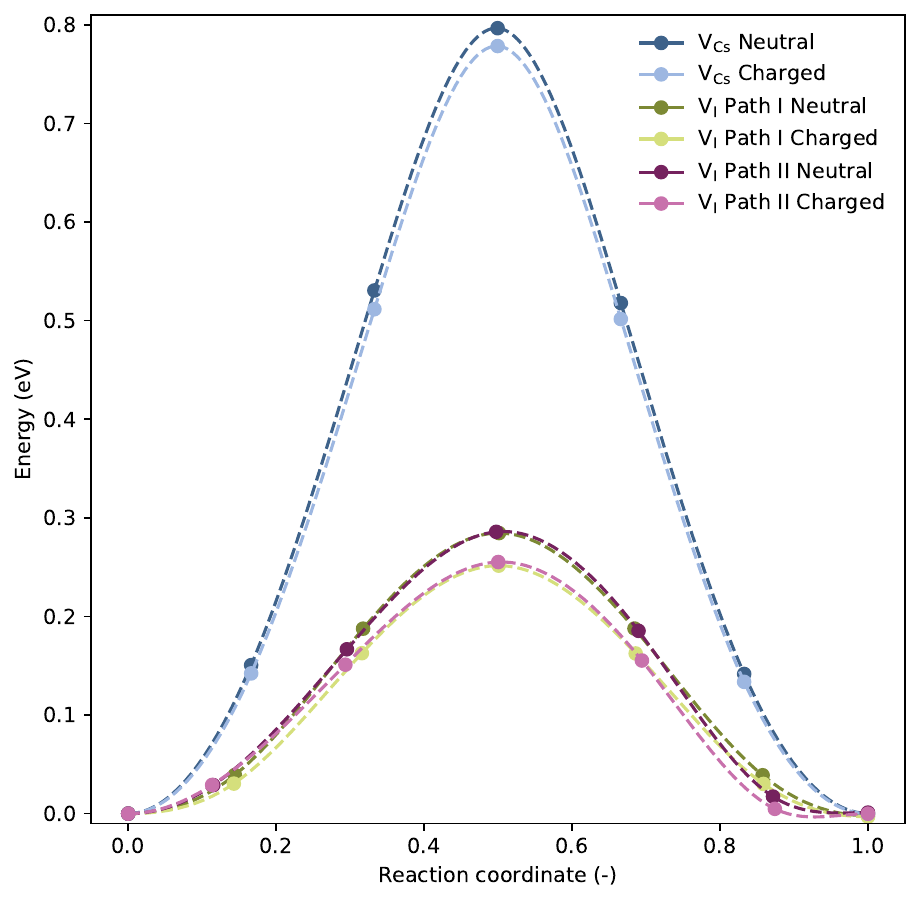}
    \caption{Migration barriers of V$_{\text{Cs}}$ and V$_{\text{I}}$ in bulk \ce{CsGeI3} in charged and neutral cells calculated using the PBEsol functional.}
    \label{fig:defect_mig_CsGeI3}
\end{figure}

\begin{figure}[ht]
    \centering
    \includegraphics[width=0.5\linewidth]{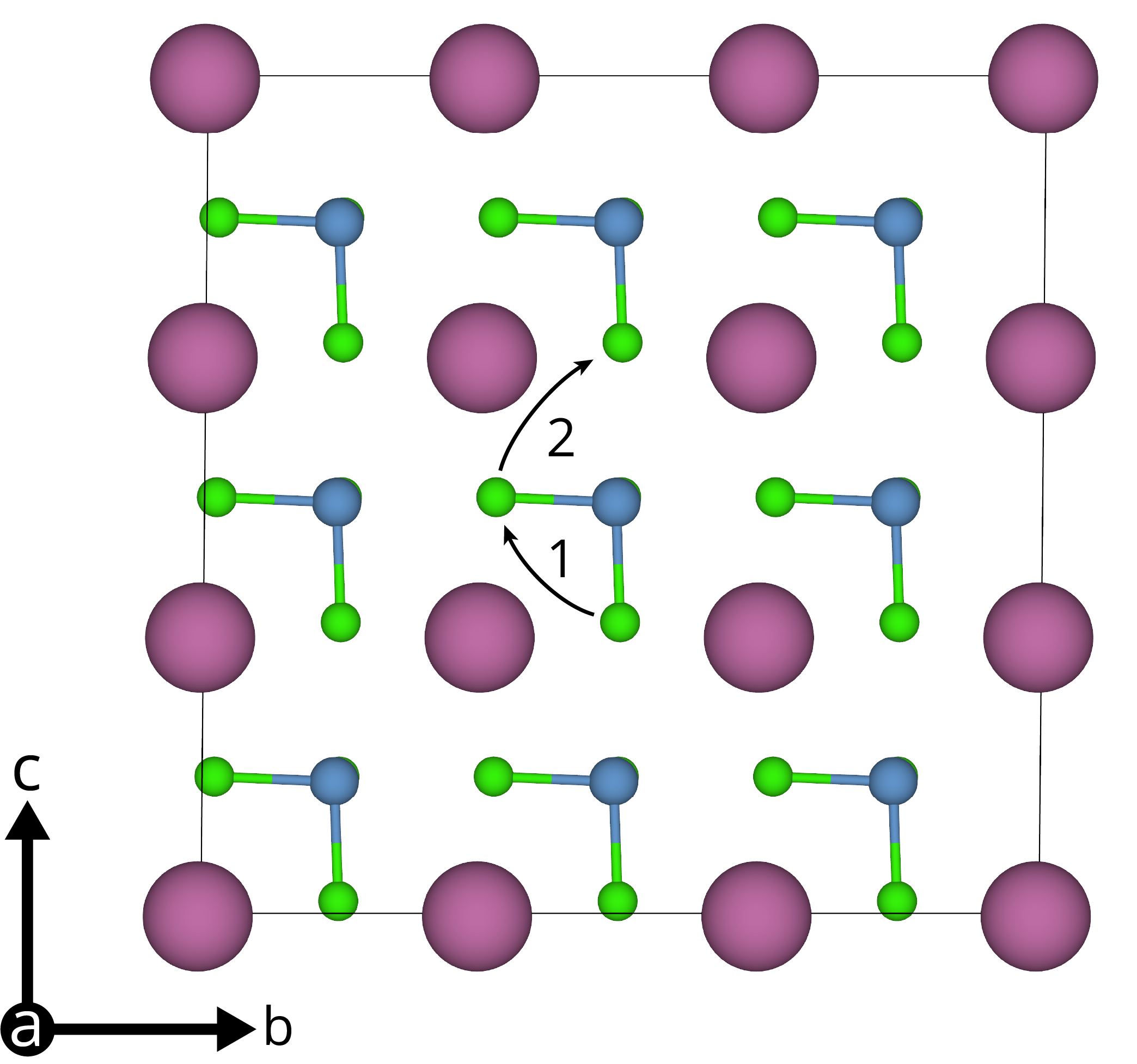}
    \caption{The two vacancy paths investigated shown in a $3\times3\times3$ supercell of \ce{CsGeCl3}. Only the three short bonds in each Ge-Cl octahedra are shown to make the difference between the two paths obvious.}
    \label{fig:vacancy_paths}
\end{figure}

\begin{figure}[ht]
    \centering
    \includegraphics[width=0.95\linewidth]{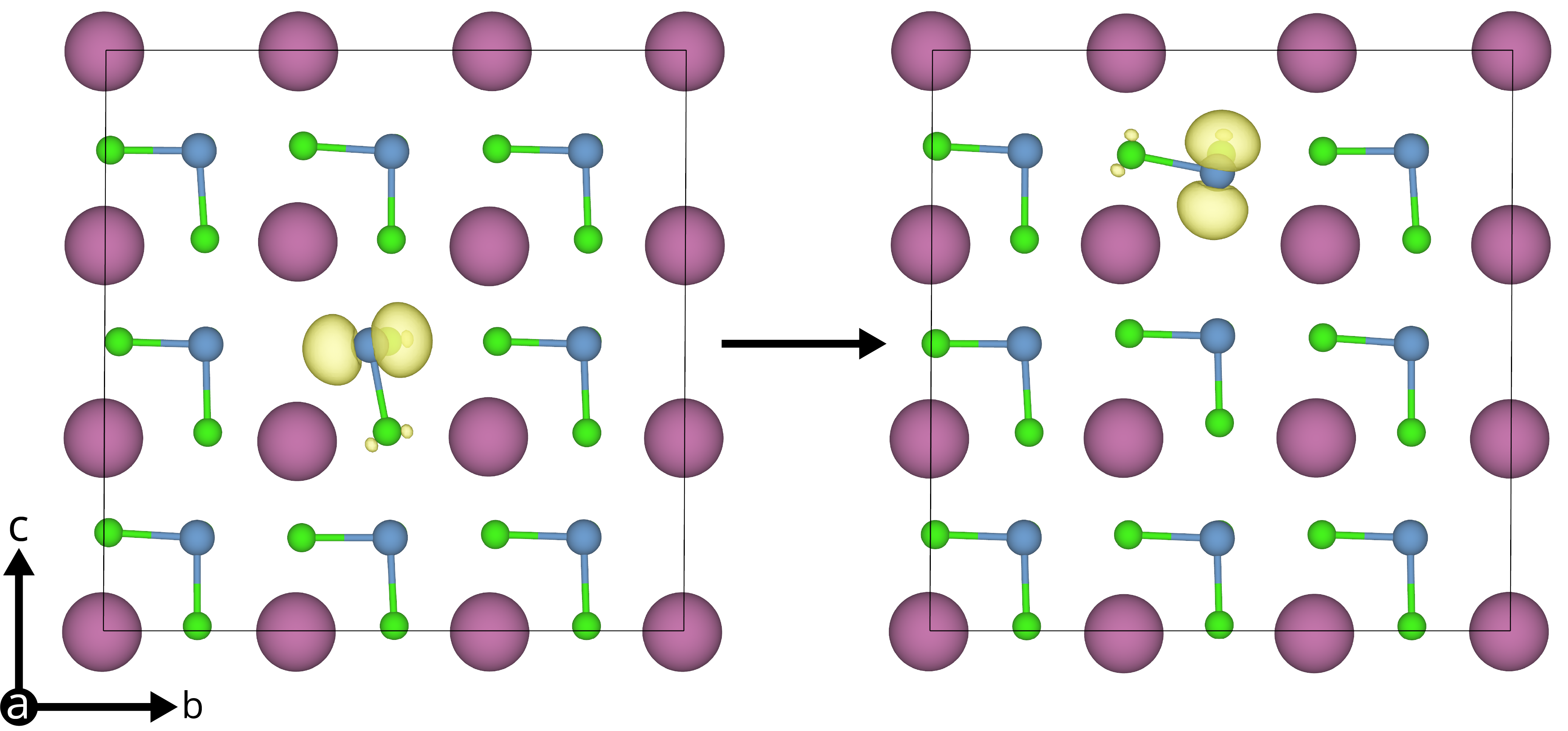}
    \caption{Partial charge density showing the charge compensating electron moving from one Ge atom to another Ge atom as a result of vacancy migration using path II.}
    \label{fig:parchg}
\end{figure}

Calculated migration barriers for V$_{\text{Cs}}$ and V$_{\text{I}}$ in \ce{CsGeI3}, displayed in SI Figure \ref{fig:defect_mig_CsGeI3}, show very similar results to V$_{\text{Cs}}$ and V$_{\text{Cl}}$ in \ce{CsGeCl3} (main text Figure \ref{main-fig:defect_mig}), indicating that halogen vacancies are very mobile in halide perovskites. SI Figure \ref{fig:vacancy_paths} show the two different halogen vacancy paths investigated. Path I is between two halogen sites with the shortest Ge-X bond to the same Ge atom, while path I is between two sites with the shortest Ge-X bond to two different Ge atoms. Path I also requires a transfer of an electron between two Ge atoms. The partial charge densities for the start and end structures of migration using path I are displayed in SI Figure \ref{fig:parchg}.

\clearpage

\section{Domain Walls}

\begin{figure}[ht]
    \centering
    \includegraphics[width=0.95\linewidth]{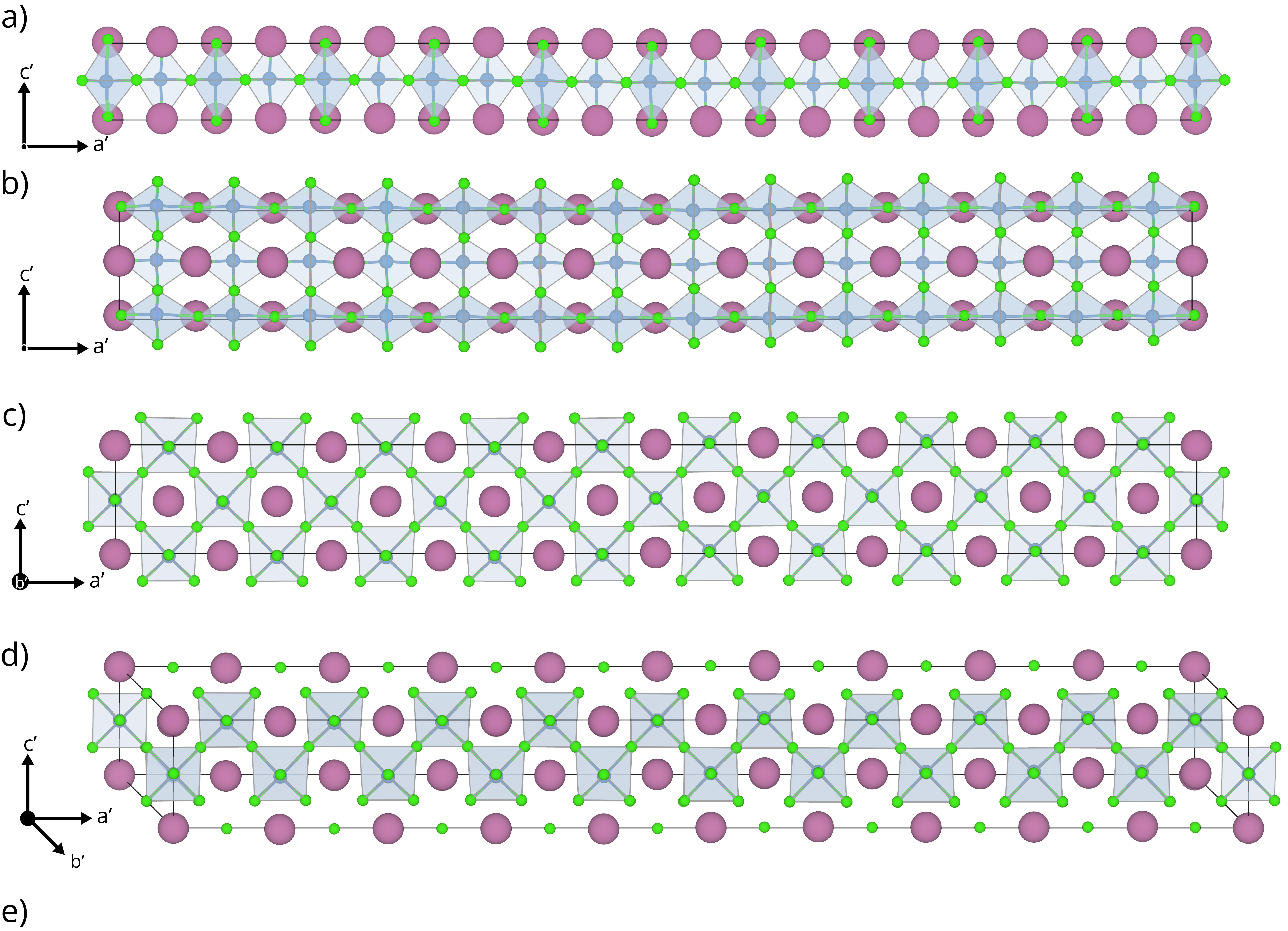}
    \caption{71$\degree$ (a), 109$\degree$ (b), and 180$\degree$ Y- (c) and X-type (d) DW cells for \ce{CsGeCl3} optimised using the PBEsol functional.}
    \label{fig:DW_cells}
\end{figure}

Table SI \ref{tab:DW_lattice_calc} shows how the lattice parameters for 71$\degree$, 109$\degree$, and 180$\degree$ DW supercells can be calculated. $a$ and $\alpha$ are the lattice parameters from the primitive cell and $n$ determines the number of perovskites units in the supercell. Examples of the different DW supercells are displayed in SI Figure \ref{fig:DW_cells}. The ferroelastic DWs required for the 71$\degree$ and 109$\degree$ DWs in \ce{CsGeCl3} are not very pronounced, as the rhombohedral angle is very close to 90$\degree$.

\renewcommand{\arraystretch}{2}
\begin{table}[ht]
    \caption{Formulas for calculating lattice parameters from primitive lattice parameters for 71$\degree$, 109$\degree$, and 180$\degree$ Y- and X-type DW supercells.}
    \centering
    \begin{tabular}{ccccccc}
         & $a'$ & $b'$ & $c'$ & $\alpha'$ & $\beta'$ & $\gamma'$  \\
        \toprule
        $\begin{array}{c} \displaystyle 71^{\circ} \end{array}$ & $\begin{array}{c} \displaystyle n\cdot a \cos{\frac{\alpha}{2}} \end{array}$ & $\begin{array}{c} \displaystyle 2a\sin{\frac{\alpha}{2}} \end{array}$ & $\begin{array}{c} \displaystyle a\end{array}$ & $\begin{array}{c} \displaystyle 90^{\circ} \end{array}$ & $\begin{array}{c} \displaystyle 90^{\circ} \end{array}$ & $\begin{array}{c} \displaystyle 90^{\circ} \end{array}$ \\ 
        
        $\begin{array}{c} \displaystyle 109^{\circ} \end{array}$ & $\begin{array}{c} \displaystyle n\cdot a\sqrt{1-\left(\frac{\cos{\alpha}}{\cos{\frac{\alpha}{2}}}\right)^2} \end{array}$ & $\begin{array}{c} \displaystyle 2a\sin{\frac{\alpha}{2}} \end{array}$ & $\begin{array}{c} \displaystyle 2a\cos{\frac{\alpha}{2}} \end{array}$ & $\begin{array}{c} \displaystyle 90^{\circ} \end{array}$ & $\begin{array}{c} \displaystyle 90^{\circ} \end{array}$ & $\begin{array}{c} \displaystyle 90^{\circ} \end{array}$ \\ 
        
        $\begin{array}{c} \displaystyle 180^{\circ} \,\text{Y} \end{array}$ & $\begin{array}{c} \displaystyle n\cdot a\sin{\frac{\alpha}{2}}\end{array}$ & $\begin{array}{c} \displaystyle a \end{array}$ & $\begin{array}{c} \displaystyle 2a\cos{\frac{\alpha}{2}} \end{array}$ & $\begin{array}{c} \displaystyle \arccos{\frac{\cos{\alpha}}{\cos{\frac{\alpha}{2}}}} \end{array}$ & $\begin{array}{c} \displaystyle 90^{\circ} \end{array}$ & $\begin{array}{c} \displaystyle 90^{\circ} \end{array}$ \\ 

        $\begin{array}{c} \displaystyle 180^{\circ} \,\text{X} \end{array}$ & $\begin{array}{c} \displaystyle n\cdot \frac{a}{2}\cos{\frac{\alpha}{2}}\end{array}$ & $\begin{array}{c} \displaystyle 2a\cos{\frac{\alpha}{2}} \end{array}$ & $\begin{array}{c} \displaystyle a\sqrt{3+6\cos{\alpha}} \end{array}$ & $\begin{array}{c} \displaystyle 90^{\circ} \end{array}$ & $\begin{array}{c} \displaystyle 90^{\circ} \end{array}$ & $\begin{array}{c} \displaystyle 120^{\circ} \end{array}$ \\ 
        \midrule
    \end{tabular}

    \label{tab:DW_lattice_calc}
\end{table}

\clearpage

\begin{figure}[ht]
    \centering
    {\includegraphics[width=0.45\linewidth]{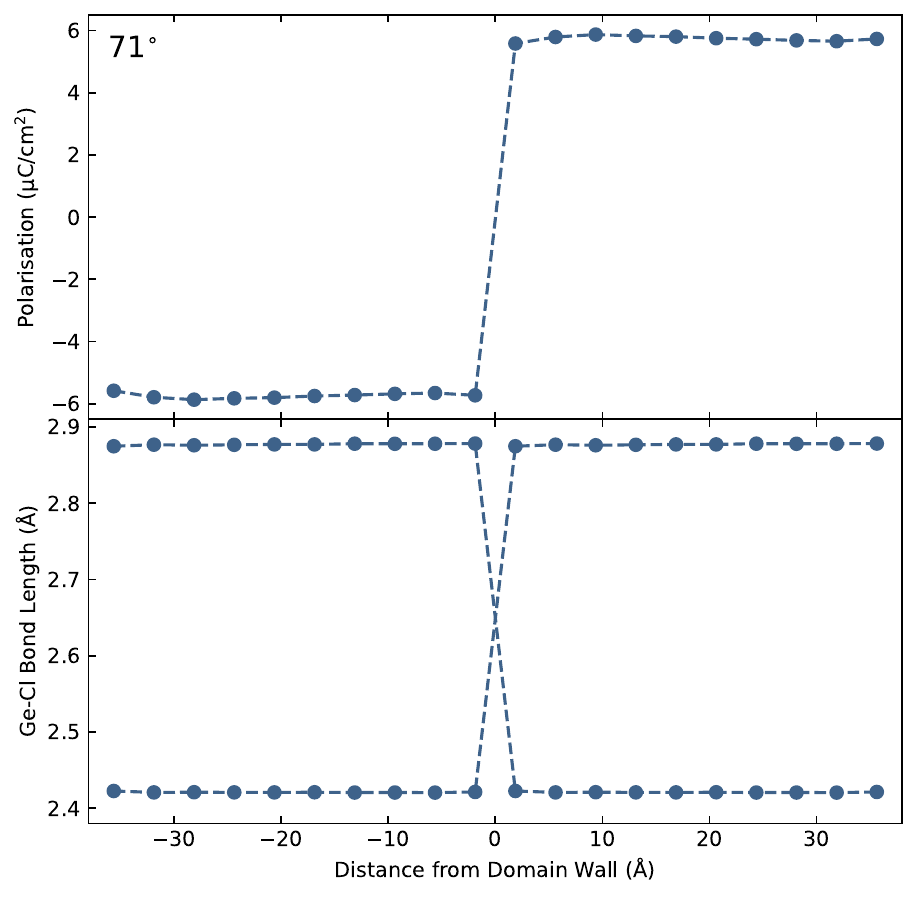}}
    \quad
    {\includegraphics[width=0.45\linewidth]{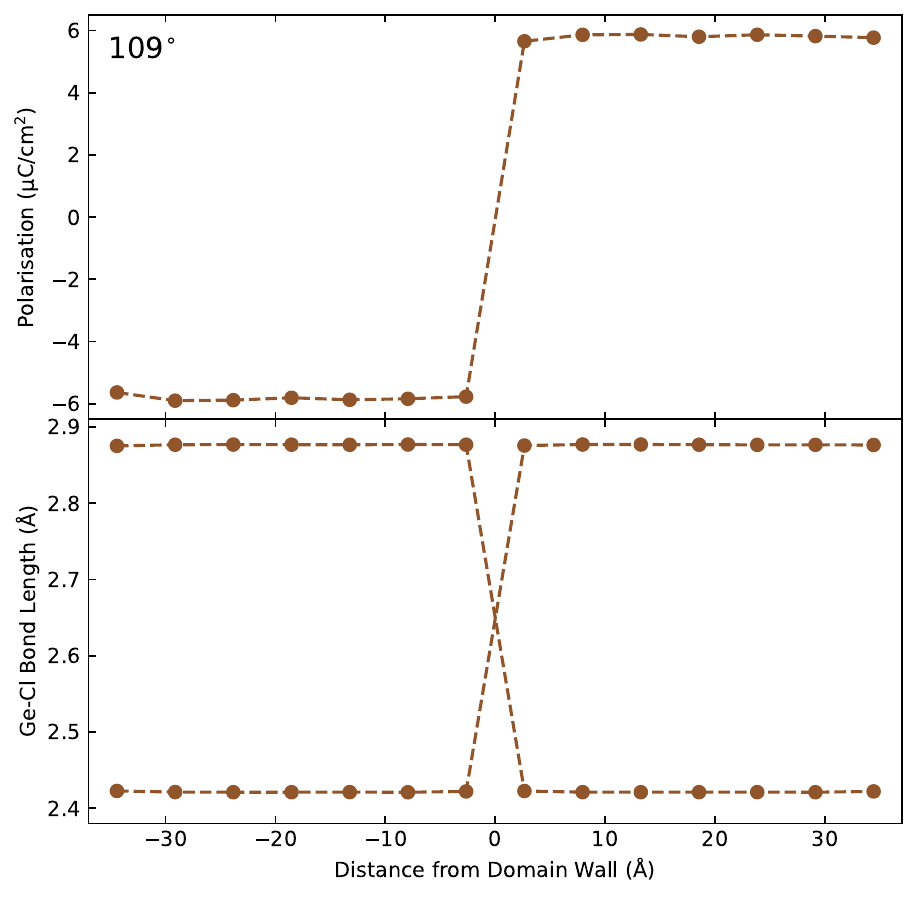}}
    \quad
    {\includegraphics[width=0.45\linewidth]{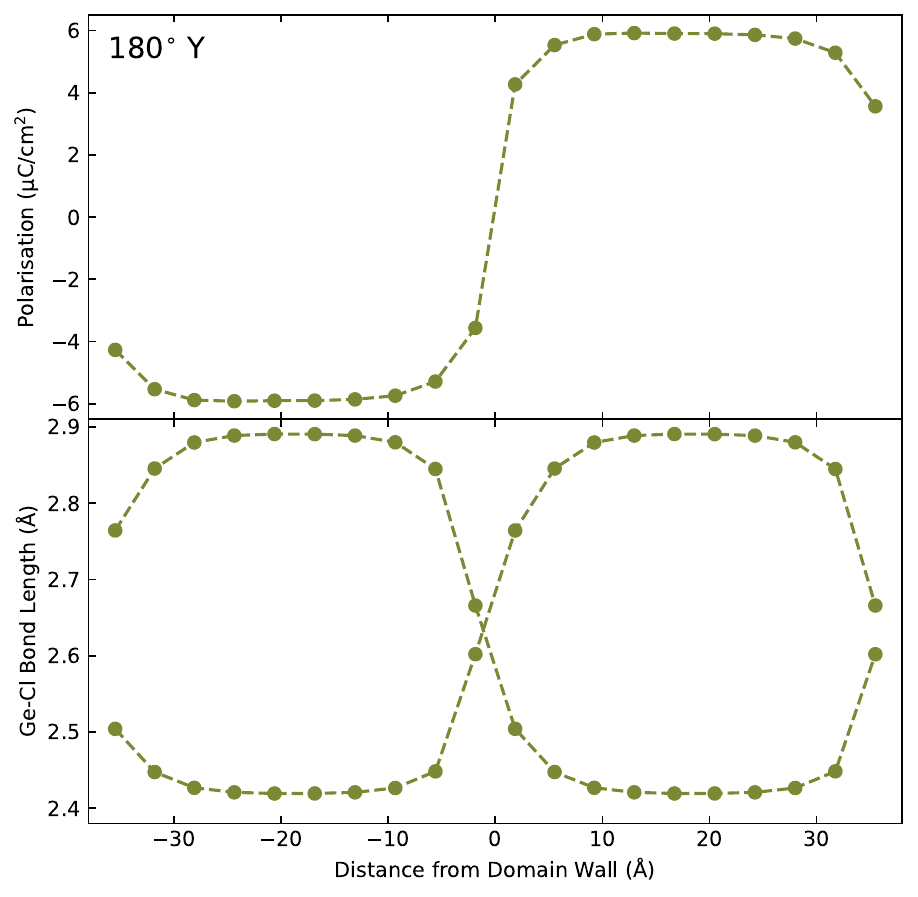}}
    \quad
    {\includegraphics[width=0.45\linewidth]{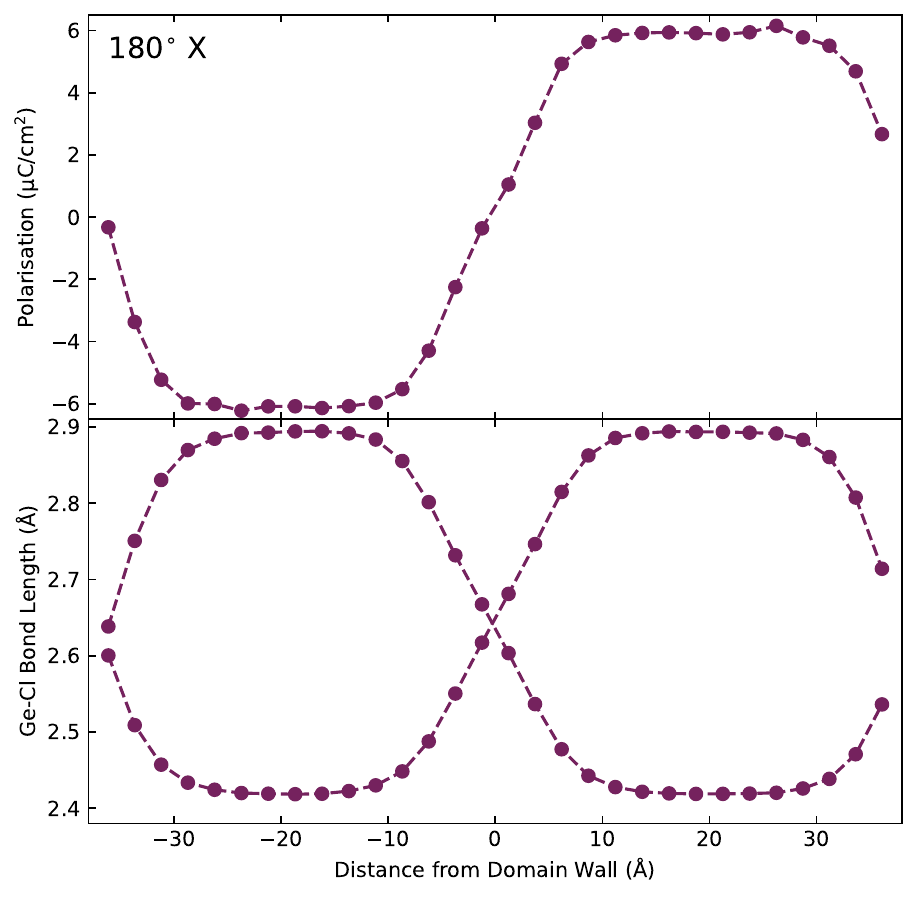}}
    \caption{Polarisation and Ge-Cl distances as a function of the DW cell for 71$\degree$, 109$\degree$, and 180$\degree$ Y- and X-type DWs in \ce{CsGeCl3}.}
    \label{fig:dw_para}
\end{figure}

\begin{figure}[ht]
    \centering
    {\includegraphics[width=0.45\linewidth]{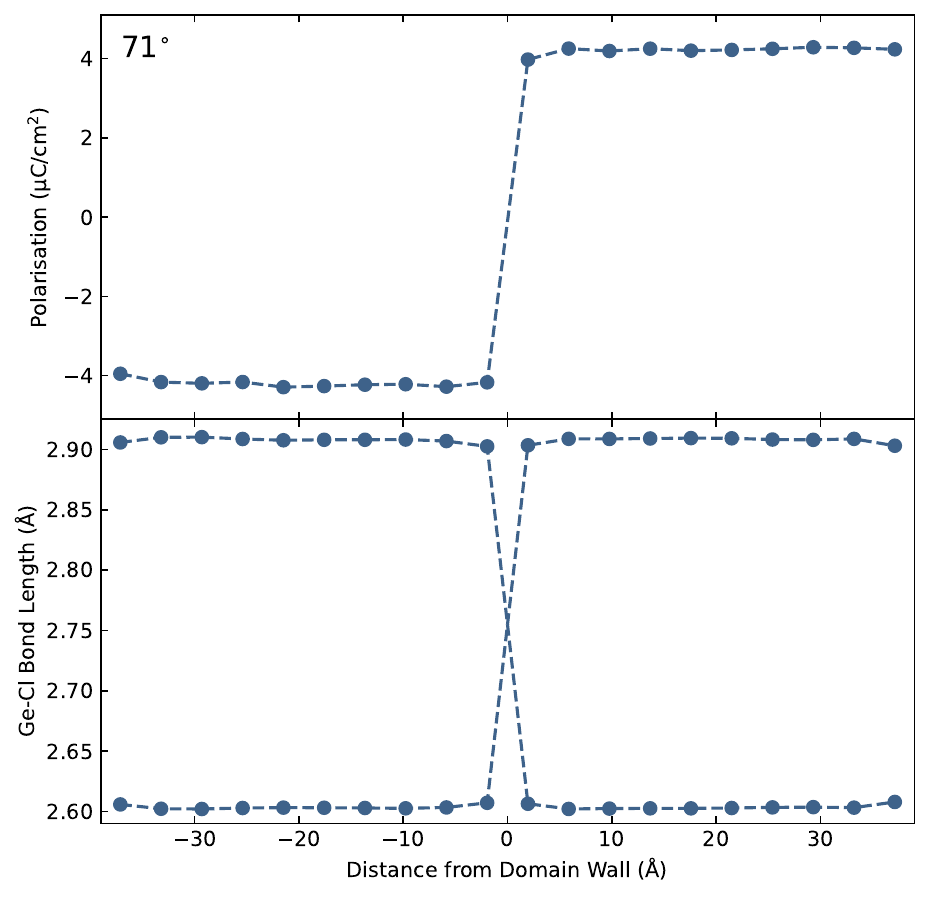}}
    \quad
    {\includegraphics[width=0.45\linewidth]{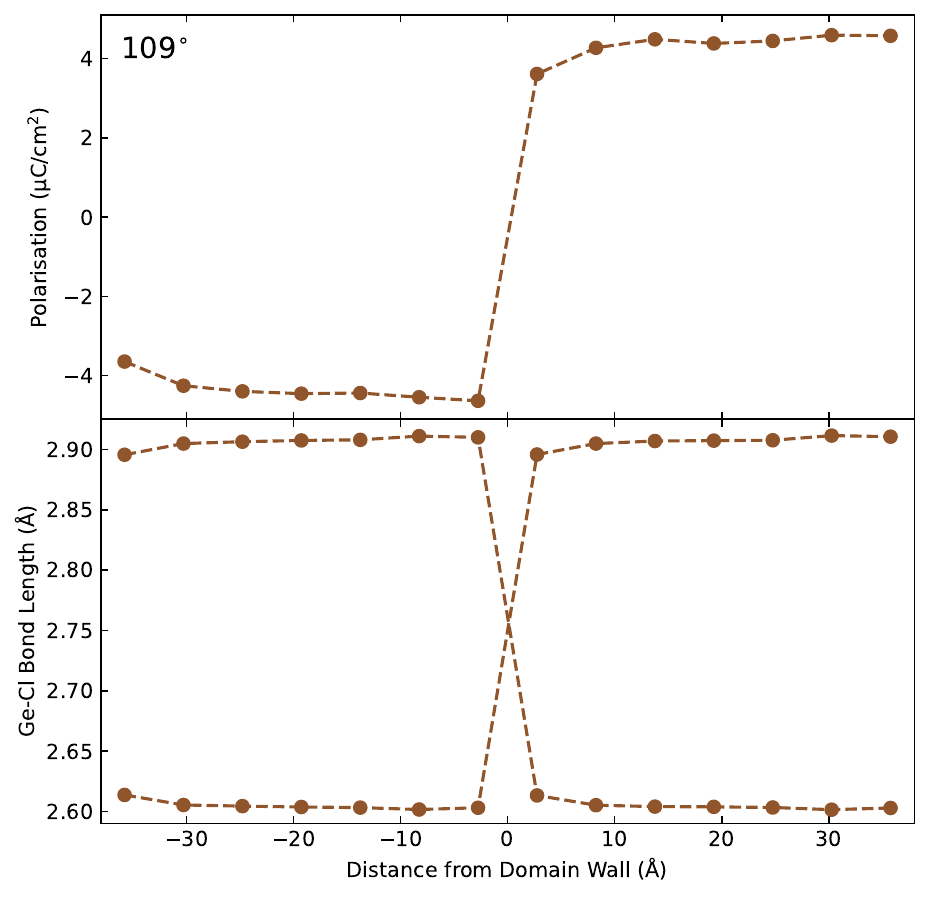}}
    \quad
    {\includegraphics[width=0.45\linewidth]{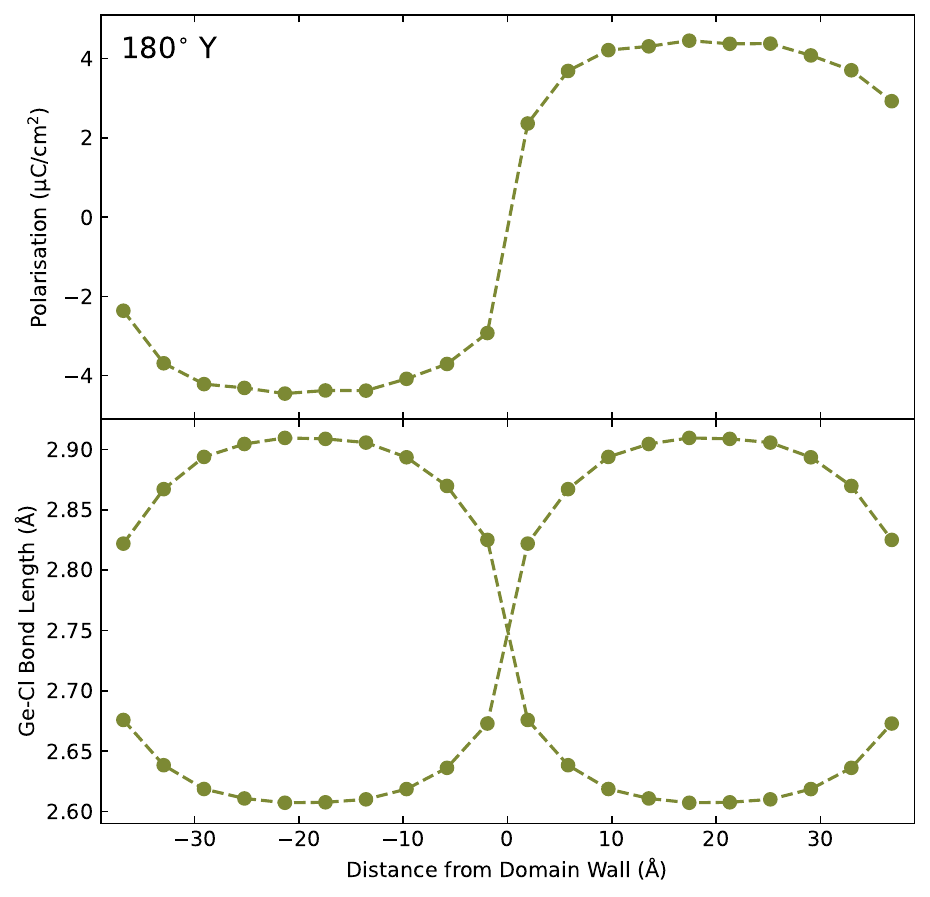}}
    \quad
    {\includegraphics[width=0.45\linewidth]{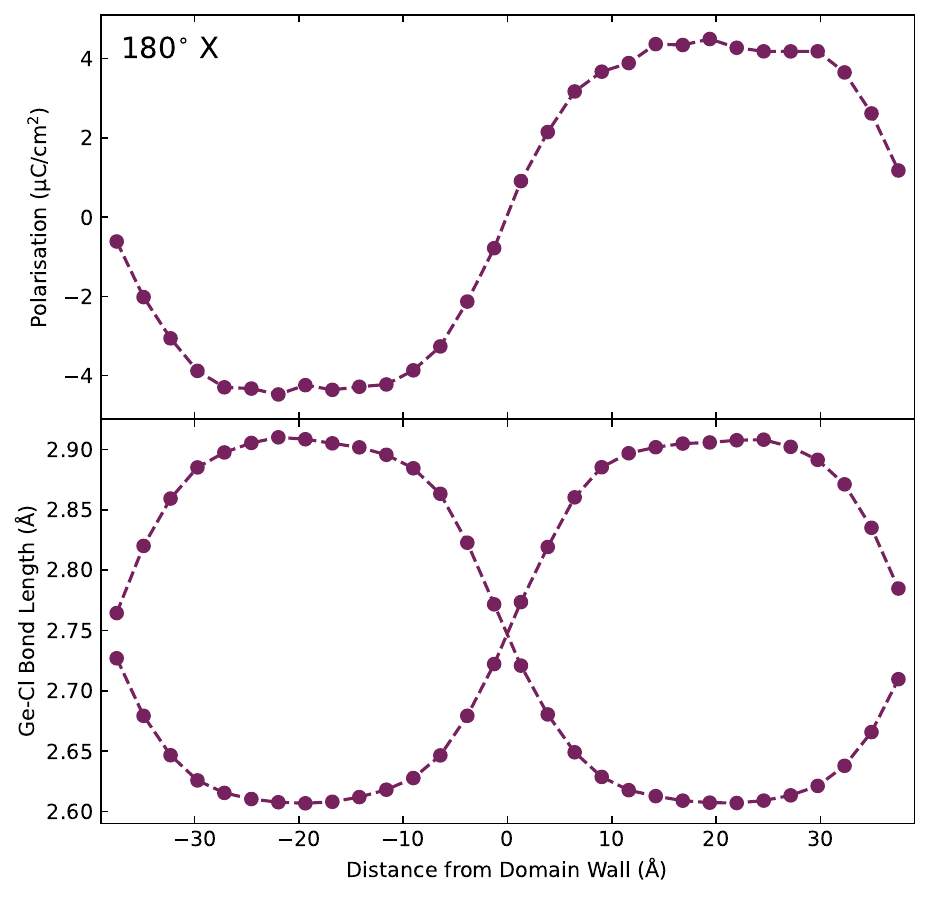}}
    \caption{Polarisation and Ge-Br distances as a function of the DW cell for 71$\degree$, 109$\degree$, and 180$\degree$ Y- and X-type DWs in \ce{CsGeBr3}.}
    \label{fig:dw_para_CsGeBr3}
\end{figure}

\begin{figure}[ht]
    \centering
    {\includegraphics[width=0.45\linewidth]{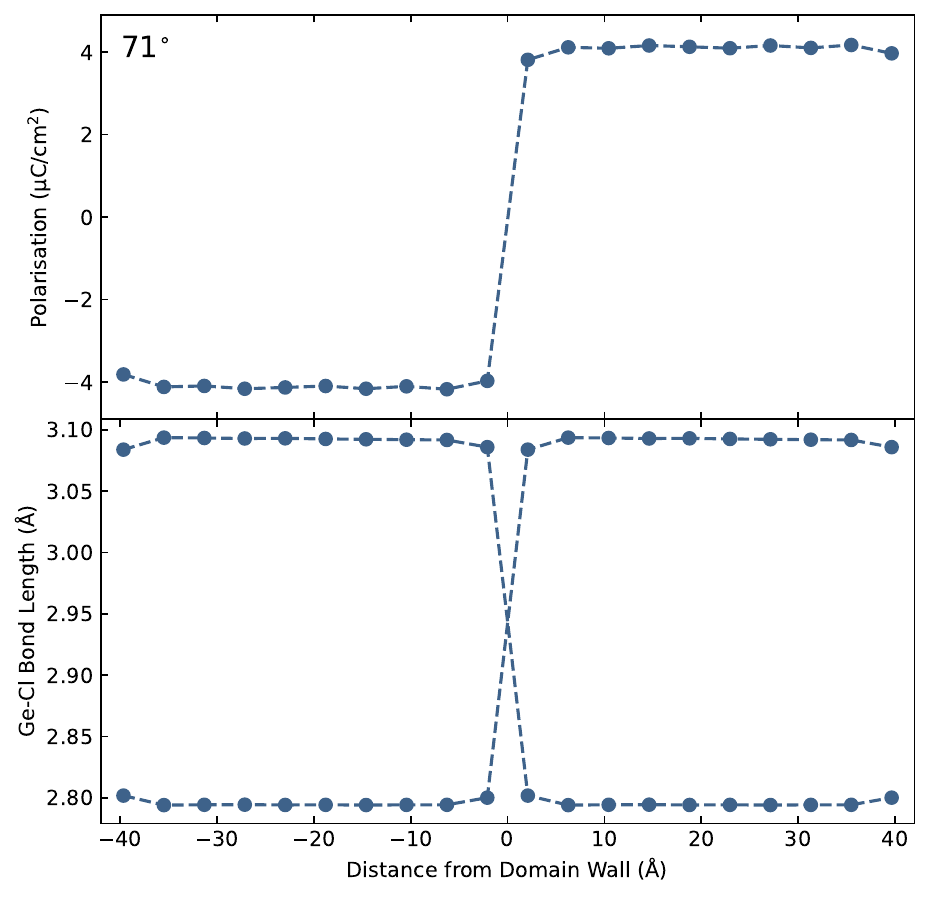}}
    \quad
    {\includegraphics[width=0.45\linewidth]{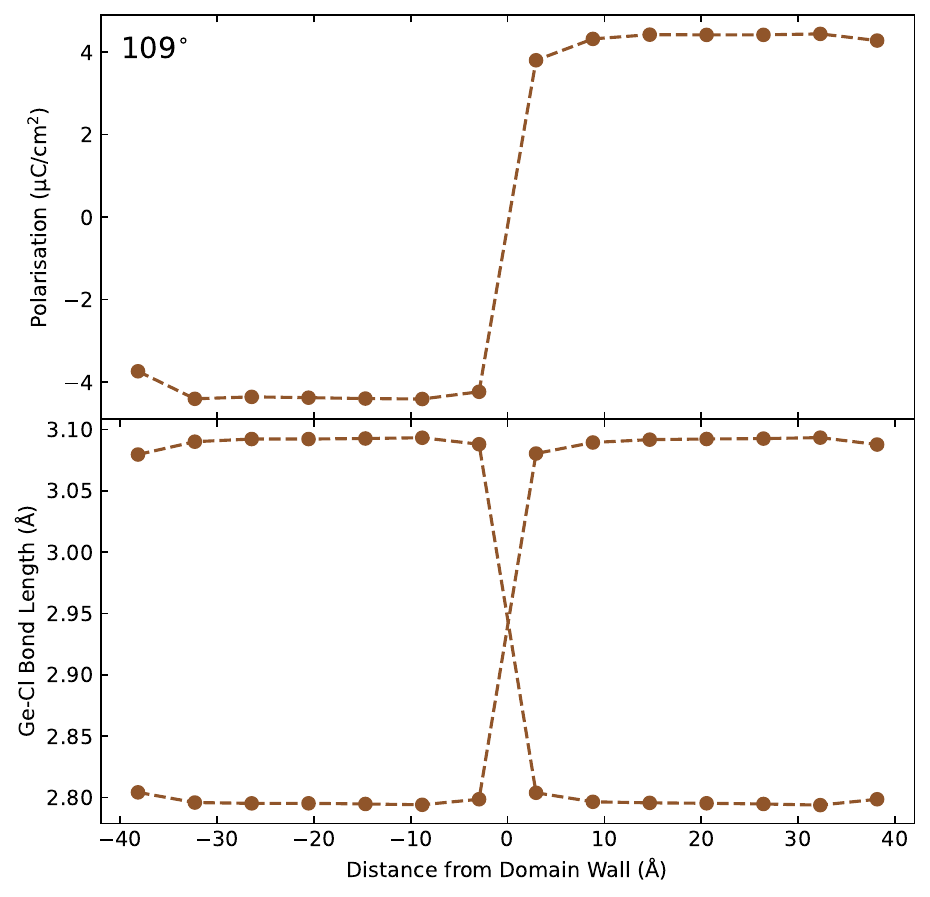}}
    \quad
    {\includegraphics[width=0.45\linewidth]{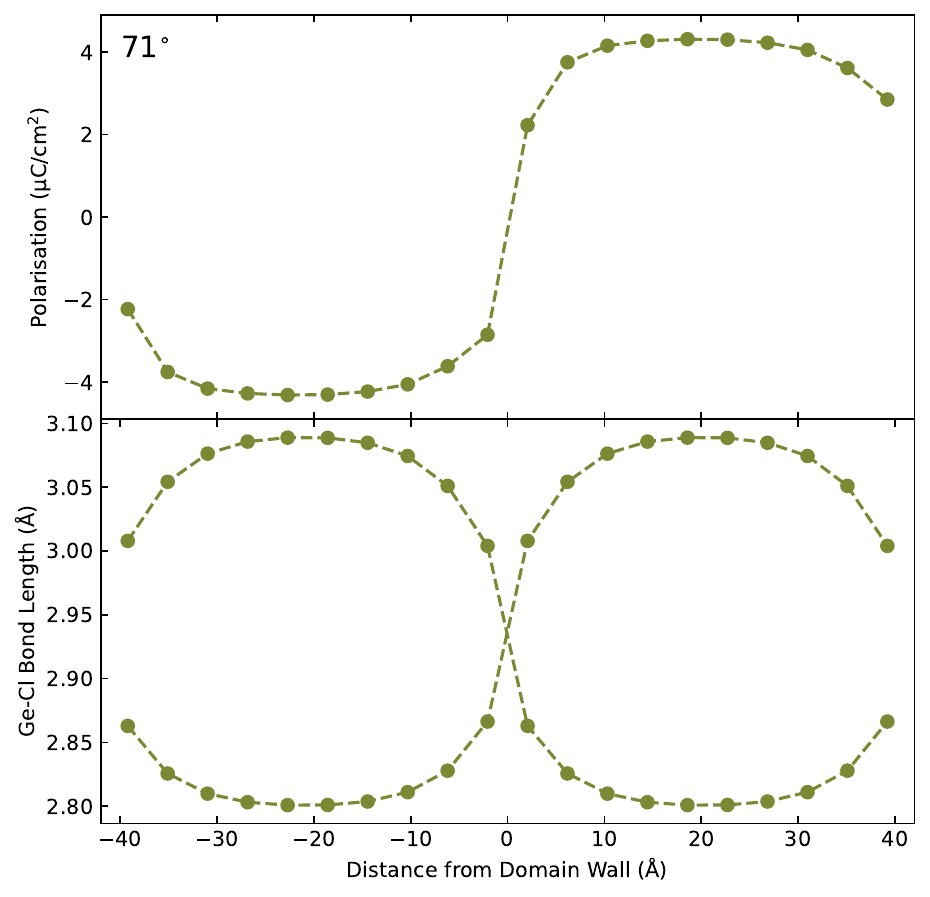}}
    \quad
    {\includegraphics[width=0.45\linewidth]{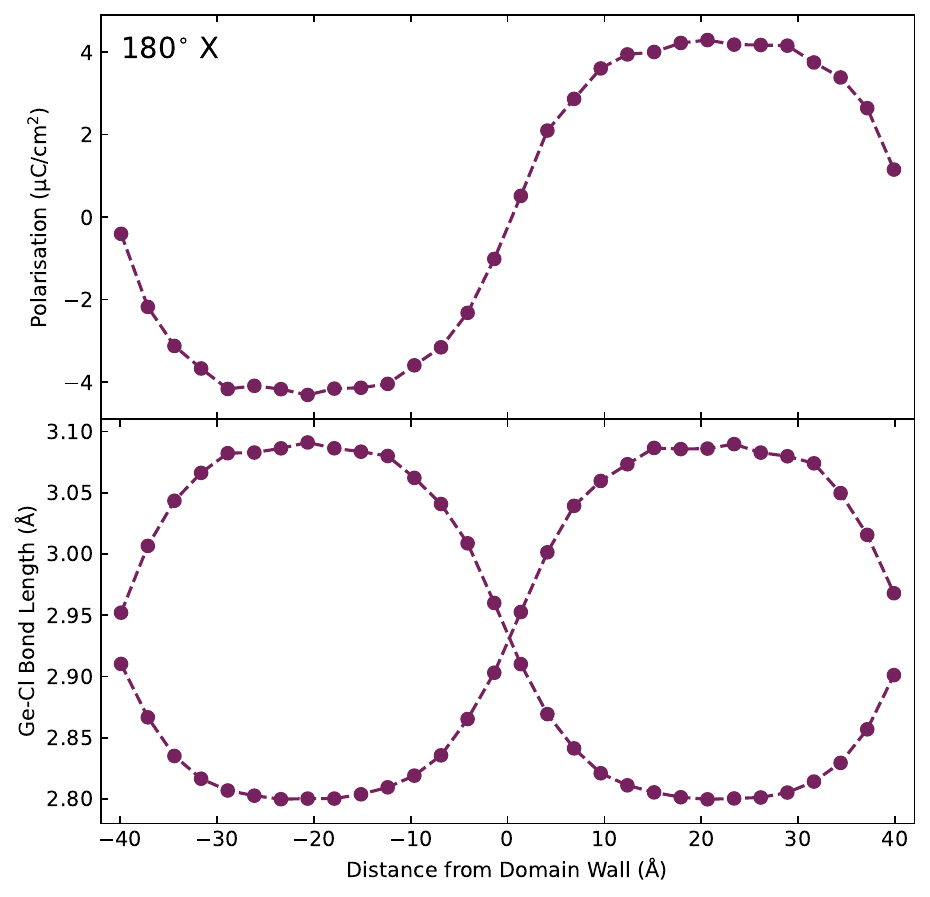}}
    \caption{Polarisation and Ge-I distances as a function of the DW cell for 71$\degree$, 109$\degree$, and 180$\degree$ Y- and X-type DWs in \ce{CsGeI3}.}
    \label{fig:dw_para_CsGeI3}
\end{figure}

Polarisation, calculated from formal charges, and Ge-X bond lengths as a function of distance from the DW in relaxed supercells of \ce{CsGeCl3}, \ce{CsGeBr3} and \ce{CsGeI3} are shown in SI Figure \ref{fig:dw_para}, \ref{fig:dw_para_CsGeBr3} and \ref{fig:dw_para_CsGeI3}, respectively.

\clearpage

\begin{figure}[ht]
    \centering
    \includegraphics[width=0.9\linewidth]{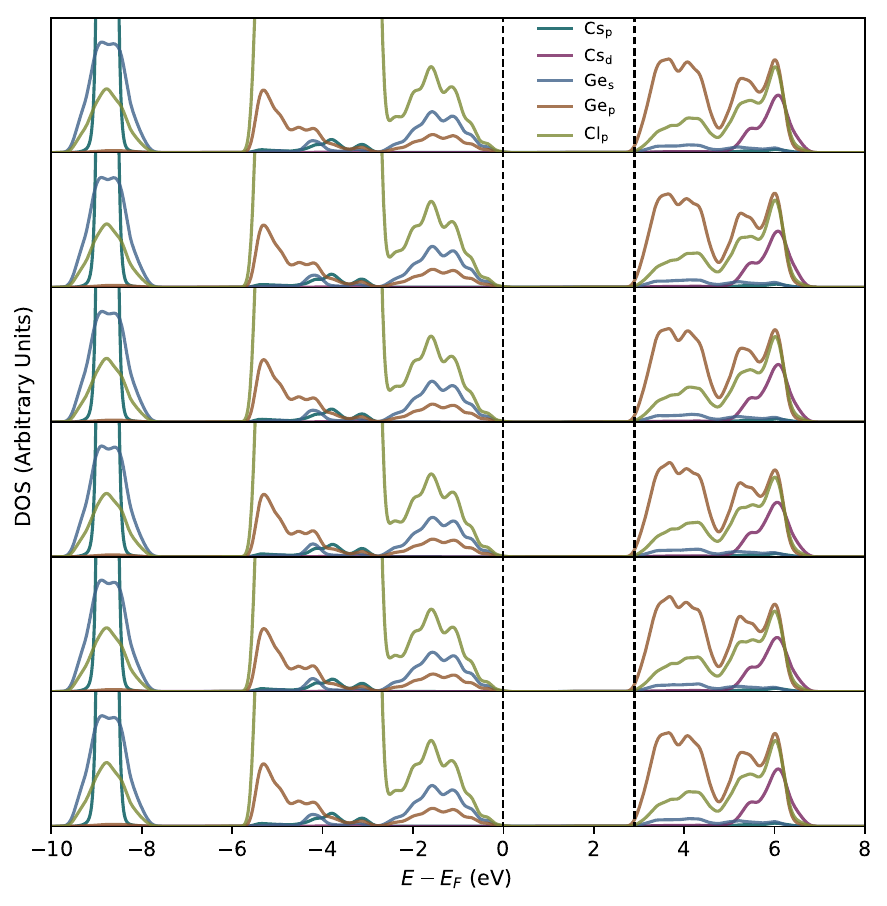}
    \caption{Layer-resolved electronic density of states for a 71$\degree$ domain wall calculated using the HSE06 functional.}
    \label{fig:layer_resolved}
\end{figure}

SI Figure \ref{fig:layer_resolved} shows the electronic DOS as a function of the DW cell for a 71$\degree$ in \ce{CsGeCl3}. No visible changes to the DOS at and around the DW can be observed.

\clearpage

\begin{figure}[ht]
    \centering
    \includegraphics[width=0.85\linewidth]{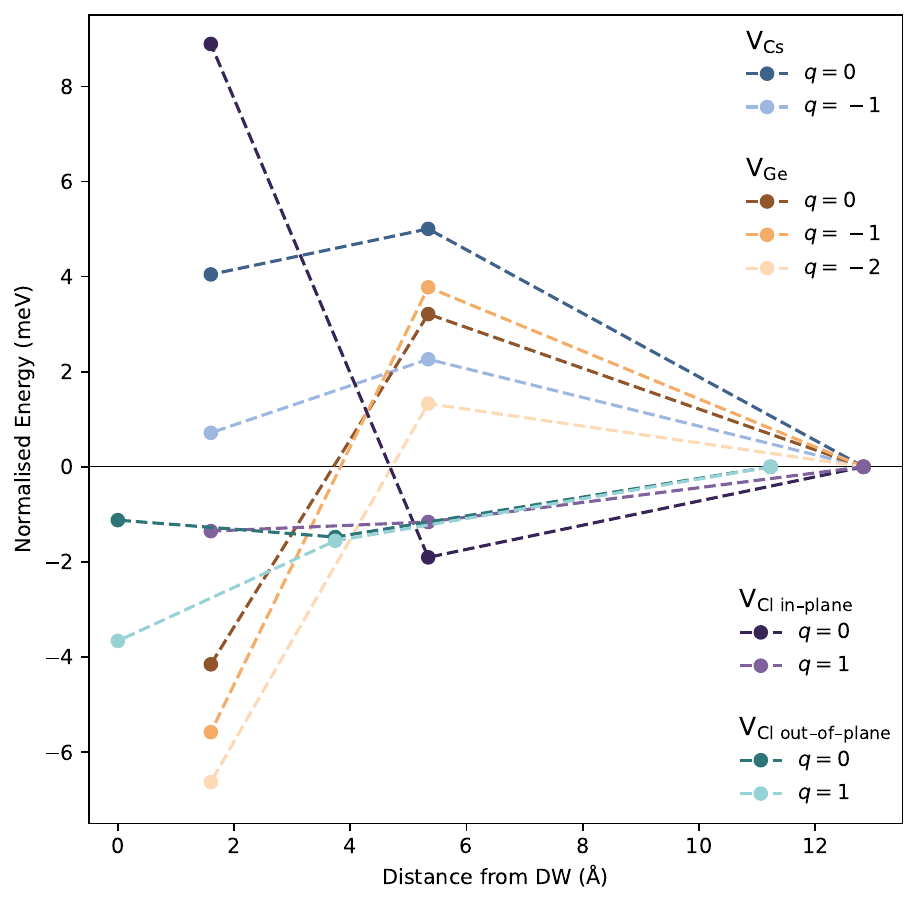}
    \caption{Normalised defect formation energies as a function of distance from a 71$\degree$ DW.}
    \label{fig:defect_dw}
\end{figure}

Normalised defect formation energies as a function of the distance to a DW are shown in SI Figure \ref{fig:defect_dw}. The formation energies are normalised to the defect formation energy of the defect located in the middle of a domain. Most defects show a slight reduction in energy at the DW, however, the energy differences are tiny and therefore do not indicate any significant accumulation of defects at these DWs in \ce{CsGeCl3}.

\clearpage

\begin{figure}[ht]
    \centering
    \includegraphics[width=0.75\linewidth]{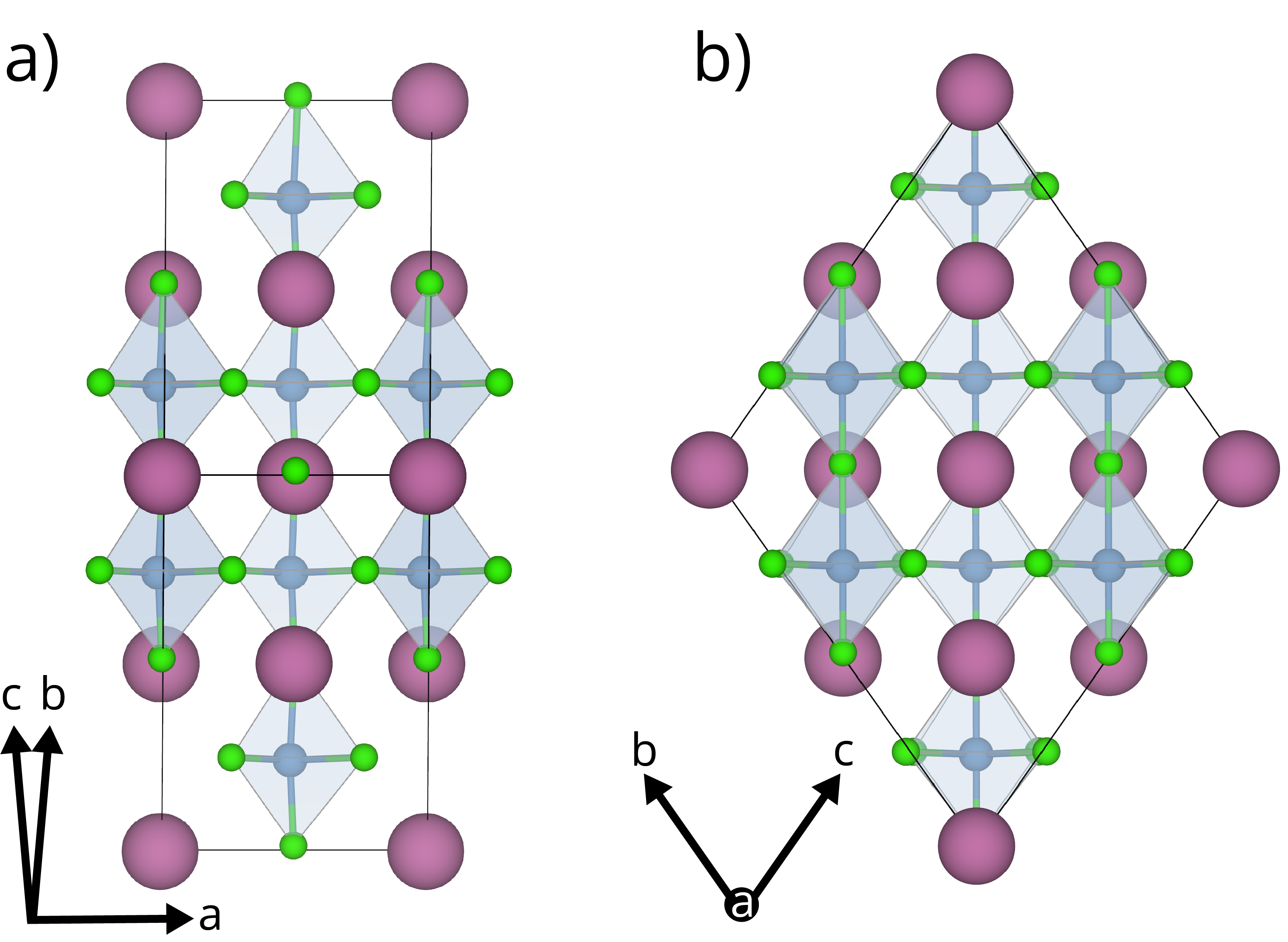}
    \caption{Visualisation of the building block ($n=1$) used to create supercells for studying DW pinning by point defects. a) and b) is the same structure viewed from different directions.}
    \label{fig:NEB_DWs}
\end{figure}

The supercells used for investigations of DW pinning by point defects were created using the equations displayed in SI Table \ref{tab:DW_NEB_lattice_calc}. A single domain unit ($n=1$) was made by a  $\left[\begin{smallmatrix} 
 1 & 0 & 1\\
 1 & 2 & -1\\
 -1 & 2 & 1
 \end{smallmatrix}\right]$
 expansion of the primitive unit cell and displayed in SI Figure \ref{fig:NEB_DWs}. As this cell does not include any ferroelastic DW, there is a very small angle between a and b$\times$c, which is not in the DW cell created from SI Table \ref{tab:DW_NEB_lattice_calc}.

\renewcommand{\arraystretch}{1.2}
\begin{table}[ht]
    \caption{Formulas for calculating lattice parameters from primitive lattice parameters for the 71$\degree$ DW supercell used for ci-NEB calculations with defects. $n$ is the number of units shown in Figure \ref{fig:NEB_DWs} included.}
    \centering
    \begin{tabular}{cccccc}
         $a'$ & $b'$ & $c'$ & $\alpha'$ & $\beta'$ & $\gamma'$  \\
        \toprule
        $\begin{array}{c} \displaystyle n\cdot a\sqrt{2(1-\cos{\alpha})(1+2\cos{\alpha})} \end{array}$ & $\begin{array}{c} \displaystyle a\sqrt{6-2\cos{\alpha}} \end{array}$ & $\begin{array}{c} \displaystyle a\sqrt{6-2\cos{\alpha}}\end{array}$ & $\begin{array}{c} \displaystyle \arccos{\frac{1+\cos{\alpha}}{3-\cos{\alpha}}} \end{array}$ & $\begin{array}{c} \displaystyle 90^{\circ} \end{array}$ & $\begin{array}{c} \displaystyle 90^{\circ} \end{array}$ \\ 
         
        \midrule
    \end{tabular}

    \label{tab:DW_NEB_lattice_calc}
\end{table}

\clearpage

\footnotesize

\bibliographystyle{unsrt}
\bibliography{references}